\documentclass{elsart}

\usepackage{graphicx,epsfig,amsmath,amssymb}
\usepackage{bm}
\usepackage{lscape}
\usepackage{dcolumn}
\usepackage{amssymb}
\usepackage{longtable}

\begin{document}

\begin{frontmatter}



\title{Perturbed angular correlations for Gd in gadolinium: in-beam
comparisons of relative magnetizations}

\author[label1]{A.E. Stuchbery\corauthref{1}},
\author[label1,label2]{A.N. Wilson},
\author[label1]{P.M. Davidson},
\author[label3]{N. Benczer-Koller}

\address[label1]{Department of Nuclear Physics, Research School of Physical
Sciences and Engineering, The Australian National University,
Canberra, ACT 0200, Australia}
\address[label2]{Department of Physics, The Australian National University,
Canberra, ACT 0200, Australia}
\address[label3]{Department of Physics and Astronomy, Rutgers University, New
Brunswick, New Jersey, 08903, USA}

\corauth[1]{andrew.stuchbery@anu.edu.au \ Phone: + 61 2 6125 2097
Fax: + 61 2 6125 0748}


\begin{abstract}
Perturbed angular correlations were measured for Gd ions implanted
into gadolinium foils following Coulomb excitation with 40~MeV
$^{16}$O beams. A technique for measuring the relative
magnetizations of ferromagnetic gadolinium hosts under in-beam
conditions is described and discussed. The combined
electric-quadrupole and magnetic-dipole interaction is evaluated.
The effect of nuclei implanted onto damaged or non-substitutional
sites is assessed, as is the effect of misalignment between the
internal hyperfine field and the external polarizing field. Thermal
effects due to beam heating are discussed.
\end{abstract}


\end{frontmatter}

\section{Introduction}
\protect \label{sect:intro}
Gadolinium foils are used extensively for in-beam measurements of
hyperfine interactions and nuclear moments
\cite{ska72,bau75,kal77,rao79,ben80,hau83,hau84,via87,stu91,cub93,stu95a,rob99,spe02}.
Magnetized gadolinium foils are used in preference to iron foils in
many applications of the transient-field technique
\cite{ben80,spe02} to measure nuclear $g$~factors because larger
perturbations of the particle-$\gamma$ correlations can be obtained
under otherwise similar experimental conditions \cite{hau83,hau84}.
One disadvantage of gadolinium hosts, however, is that the
magnetization is not well controlled, being sensitive to the
crystalline structure of the foil, and varying considerably with
both the applied field and the temperature, even at temperatures
well below the Curie temperature of 293 K \cite{nigh}.

As cooling with liquid nitrogen (77~K) is convenient in accelerator
laboratories, most in-beam measurements which employ gadolinium
hosts have been performed at somewhat higher temperatures near 90~K
because of thermal losses and the effects of beam heating on the
target. In transient-field measurements an external magnetic field,
with strength typically in the range from 0.05 to 0.1 T, is applied
to polarize the gadolinium foil. Higher fields are avoided to ensure
that bending of the primary beam is negligible \cite{gol78,spe02}.
The spacial profile of the polarizing field along the beam direction
is designed to minimize beam bending effects rather than to produce
a uniform field at the target location. Depending upon the design of
the pole pieces and the location of the target, the profile of the
external field across the target may not be uniform. Under these
conditions the magnetization in the beam spot must be carefully
related to the off-line magnetization measurements.

In this paper a method of determining the relative magnetization of
gadolinium foils under in-beam conditions is described. The static
hyperfine magnetic field, which acts on the nuclei of Coulomb
excited 2$^+_1$ states in $^{154,156,158,160}$Gd, is used to probe
the local magnetization at the beam spot under in-beam conditions.

Measurements similar to the present work were performed by Skaali
{\em et al.} \cite{ska72}, and Kalish {\em et al.} \cite{kal77}.
Additional complementary information was also obtained by H\"ausser
{\em et al.} \cite{hau83}. In these previous works, however, the
focus was on the precessions of the 4$^+_1$ and 6$^+_1$ states as
probes of the strength of the transient hyperfine magnetic field.
Here the emphasis is on the 2$^+_1$ states and the strength of the
static hyperfine magnetic field.

While the interpretation of the hyperfine interactions in terms of
the relative magnetization of different samples turns out to be
rather straight-forward, there are several additional phenomena
associated with the hyperfine fields and the ion-implantation
process that may have bearing on the interpretation of the data. The
present work will therefore include discussions of:
\begin{enumerate}
\item{the presence of the electric-field gradient in the gadolinium
matrix, which means that the hyperfine interaction cannot
universally be treated as a pure magnetic interaction}
\item{the effects of those implanted nuclei which reside on damaged
or other non-substitutional sites}
\item{the magnitude of the transient hyperfine magnetic field, which
acts on the implanted ions as they slow within the host, and which
for Gd in gadolinium has the opposite sign to the static hyperfine
field}
\item{a possible misalignment between the hyperfine magnetic field
and the external polarizing field, which can affect the
interpretation of the perturbed angular correlation when the host is
not fully saturated}
\end{enumerate}

The paper is arranged as follows: The next section
(section~\ref{sect:fieldreview}) reviews previous work on the
electric and magnetic hyperfine fields experienced by Gd ions in
gadolinium. Section~\ref{sect:IMPAC} describes the in-beam
measurements, presents examples of $\gamma$-ray and particle
spectra, and summarizes the excited-state lifetimes, $g$~factors and
quadrupole moments adopted for the analysis. The off-line
magnetization measurements are presented in
section~\ref{sect:magmeas}. Section~\ref{sect:AC} concerns the
`unperturbed' angular correlations measured at room temperature. It
includes a review of the formalism, along with the results and a
discussion of the measurements, which indicate the direction of the
electric-field gradient in the gadolinium foils. The perturbed
angular correlation results are presented in section~\ref{sect:pac}.
The formalism and data analysis procedures are described, the
effects of combined electric-quadrupole and magnetic-dipole
interactions are evaluated, as are the effects of nuclei on damaged
sites. The results are discussed in Section \ref{sect:disc}.

\section{Hyperfine fields in gadolinium hosts} \label{sect:fieldreview}

Gadolinium has a hexagonal close packed (hcp) crystal structure.
Nuclei within the gadolinium matrix therefore experience a hyperfine
electric-field gradient along with the magnetic dipole interaction.
Below the Curie temperature of 293~K, gadolinium is a simple
ferromagnet with a magnetic anisotropy that has a complex dependence
on temperature. Thus the easy direction of magnetization changes as
a function of temperature, as does the magnetic hyperfine field.
M\"ossbauer studies \cite{bau75} reveal anisotropic magnetic
hyperfine interactions in gadolinium crystals, with a field of
$|B_{\rm st}| = 37.3(5)$~T at 4.2~K oriented at 28$^\circ$ to the
$c$ axis (unpolarized sample). The electric field gradient is
apparently less sensitive to temperature \cite{via87}. The measured
splitting at 4.2~K for $^{155}$Gd,  $eqQ/h = 108\pm 1$~MHz, implies
an electric field gradient of $V_{zz} = 3.44(6) \times
10^{17}$~V/cm$^2$, assuming $Q = 1.30(2)$~b. This electric field
gradient is consistent with that found for high-spin isomers in
$^{147,148}$Gd at temperatures near
400~K \cite{via87}. 

The electric-quadrupole and magnetic-dipole hyperfine fields acting
on Gd isotopes in various gadolinium samples at different
temperatures have been studied previously by many techniques (see
Refs.~\cite{bau75,rao79,via87} and references therein). In the
present work the M\"ossbauer values at 4.2~K will be taken as the
point of reference. The sign of the static magnetic field has been
determined to be negative, for example from in-beam perturbed
angular correlation measurements \cite{ska72,kal77,hau83}.

Since the present work concerns in-beam implantation of Gd into
gadolinium, the transient hyperfine magnetic field also acts on the
ions before they come to rest. Thus the net perturbation of the
nuclear spin distribution has contributions from both the transient
and static hyperfine magnetic fields. For states of spin 4$^+$ and
higher in the ground-state band, the electric quadrupole interaction
can be ignored (see below and \cite{hau83}) and the perturbation of
the angular correlation is manifested essentially as a rotation
through the angle
\begin{equation}\label{eq:DTHETA}
\Delta \Theta = \omega \tau + \Delta \theta_{\rm tf},
\end{equation}
where $\Delta\theta_{\rm tf}$ is the precession angle due to the
transient field and $\omega\tau$ is that due to the static field.
Furthermore,
\begin{equation}
 \omega\tau = -g \,\frac{\mu_N}{\hbar}\,B_{\rm st}\,\tau ,
\end{equation}
and
\begin{equation}
 \Delta\theta_{\rm tf} = -g \,\frac{\mu_N}{\hbar} \int_{0}^{T_{s}}
B_{\rm tf}(t)\, {\rm e}^{-t / \tau}\,dt,
\end{equation}
where $\tau$ is the meanlife, $g$ the $g$~factor of the excited
nuclear state, and $T_s$ is the time taken for the recoiling ions to
stop in the ferromagnet; $B_{\rm st}$ and $B_{\rm tr}$ are the
static- and transient-field strengths, respectively. Since $B_{\rm
st}$ and $B_{\rm tr}$ have opposite signs for Gd in gadolinium, the
two contributions tend to cancel.

Equation (\ref{eq:DTHETA}) is correct only for small precession
angles because the static-field perturbation includes an attenuation
as well as the rotation of the radiation pattern. Furthermore, for
the longer-lived 2$^+_1$ states the quadrupole interaction cannot
safely be ignored. The formalism needed for a rigorous analysis of
the data is presented in section~\ref{sect:pac} along with a
discussion of the combined effect of the electric quadrupole and
magnetic dipole interactions on the observed angular correlations.

\section{IMPAC Measurements} \label{sect:IMPAC}

\subsection{Experimental Procedures} \label{sect:procedures}

Hyperfine fields acting on Gd ions implanted into gadolinium were
measured using the Implantation Perturbed Angular Correlation
(IMPAC) technique, following procedures similar to those in an
earlier study of Pt in gadolinium \cite{stu95a}. The measurements
were performed using 40~MeV $^{16}$O$^{4+}$ beams from the
Australian National University 14UD Pelletron accelerator.
Table~\ref{tab:expts} gives a summary of the angular correlation and
nuclear precession measurements performed. As will be discussed
below, the temperature shown in Table~\ref{tab:expts} is the nominal
temperature of the target frame, which does not necessarily
represent the temperature at the beam spot.

\begin{table}[t]
\caption{Summary of experiments. $T$ is the nominal temperature on
the target frame. $I_{\rm beam}$ is the beam current. Target A is
16.9 mg/cm$^2$ thick. Target B is 6.2 mg/cm$^2$ thick. }
\begin{tabular}{ccccl} \hline \hline
\multicolumn{1}{c}{Run} & \multicolumn{1}{c}{Target} &
\multicolumn{1}{c}{$T$}& \multicolumn{1}{c}{$I_{\rm beam}$} &
\multicolumn{1}{c}{Type of measurement}\\
 & & \multicolumn{1}{c}{(K)} & \multicolumn{1}{c}{(pnA)}
\\ \hline
 I   &  A   &    300  & 2.5   & $W(\theta_\gamma)$:
    $\theta_\gamma = 0^\circ,
 \pm 15^\circ, \pm 30.5^\circ, \pm 45^\circ, \pm 55^\circ, \pm 65^\circ$ \\
 II  &  A  & 90 & 2.5   & Precession: $\epsilon(\pm 65^\circ)$,
 $\epsilon(\pm 120^\circ)$\\
   & & & & $W(\theta_\gamma)$: $\theta_\gamma = 0^\circ,
 \pm 15^\circ, \pm 30.5^\circ, \pm 45^\circ, \pm 55^\circ, \pm 65^\circ$ \\
 III  &  A  & 90 & 0.375 & Precession: $\epsilon(\pm 65^\circ)$,
 $\epsilon(\pm 120^\circ)$ \\
 IV   &  B    & 90 & 2.5   & Precession: $\epsilon(\pm 65^\circ)$,
 $\epsilon(\pm 120^\circ)$\\
   & & & &$W(\theta_\gamma)$: $\theta_\gamma = \pm 31^\circ, \pm 45^\circ, \pm 65^\circ$ \\
\hline \hline
\end{tabular}
\label{tab:expts}
\end{table}

The two targets employed were the same as those used in a recent
study of transient-field strengths for high-velocity Ne and Mg ions
traversing gadolinium hosts \cite{stu05}. Target A consisted of a
rolled and annealed 16.9 mg/cm$^2$ thick gadolinium foil. To aid
with thermal conduction, the gadolinium foil was sandwiched between
two 12 $\mu$m thick indium-coated copper foils having 6~mm diameter
holes punched through at the beam position. Target B consisted of
0.1~mg/cm$^{2}$ $^{\rm nat}$C, a thin flashing of copper
(0.02~mg/cm$^{2}$) to assist adhesion, 6.2~mg/cm$^{2}$ of rolled and
annealed gadolinium, and a `thick' (5.65~mg/cm$^2$) copper backing.
Both gadolinium foils were cold rolled, beginning with 0.025~mm
thick foil of 99.9\% purity purchased from Goodfellow Cambridge
Limited. After rolling they were annealed in vacuum at $\sim
800^\circ$~C for $\sim 20$~min. To provide additional support and
thermal conduction, target B was attached to a 12~$\mu$m thick
copper foil using $\sim 1$~mg/cm$^2$ of indium as adhesive. The beam
entered the carbon side of target B. For both targets the
Coulomb-excited nuclei of the gadolinium layer recoiled with
energies of $\sim 11$~MeV and subsequently stopped within the
gadolinium layer.

Backscattered $^{16}$O ions were detected in two silicon
photodiodes, masked to expose a rectangular area 8.5~mm wide by
10.2~mm high, and placed 23.7~mm from the target 4.0~mm above and
below the beam axis at back angles. The backscattered particle
spectrum extended from $\sim 27$~MeV, due to scattering at the front
surface of the Gd foil, down to $\sim 0$ MeV due to scattering in
the depth of the target. The threshold was set at $\sim 3$~MeV. The
$^{12}$C layer of target B produced some $\alpha$-particle groups
from $^{16}$O~+~$^{12}$C reactions, which allowed an in-beam
calibration of the particle detector, but did not otherwise
interfere with the measurement. The beam species and energy were
chosen to ensure that $\gamma$~rays detected in coincidence with
backscattered beam ions originate from Coulomb-excited Gd nuclei
that recoil and stop well within the gadolinium layer.

Gamma rays were detected using two $\sim 20\%$ efficient detectors
placed $\sim 7$~cm from the target and two $\sim 50$\% efficient
high-purity Ge detectors placed 15.2~cm from the target. The larger
Ge detectors were placed at $\theta _{\gamma} = \pm 120^{\circ}$ to
the beam axis throughout the measurements. For the precession
measurements, the forward Ge detectors were placed at $\theta
_{\gamma} = \pm 65^{\circ}$, near the maximum slope of the
particle-$\gamma$ angular correlation. These detectors were also
moved through a sequence of angles to measure the angular
correlations; see Table~\ref{tab:expts}.

\subsection{Particle and $\gamma$-ray spectra} \label{sect:spectra}

\begin{figure}
    \resizebox{0.8\textwidth}{!}{
  \includegraphics[height=.8\textheight]{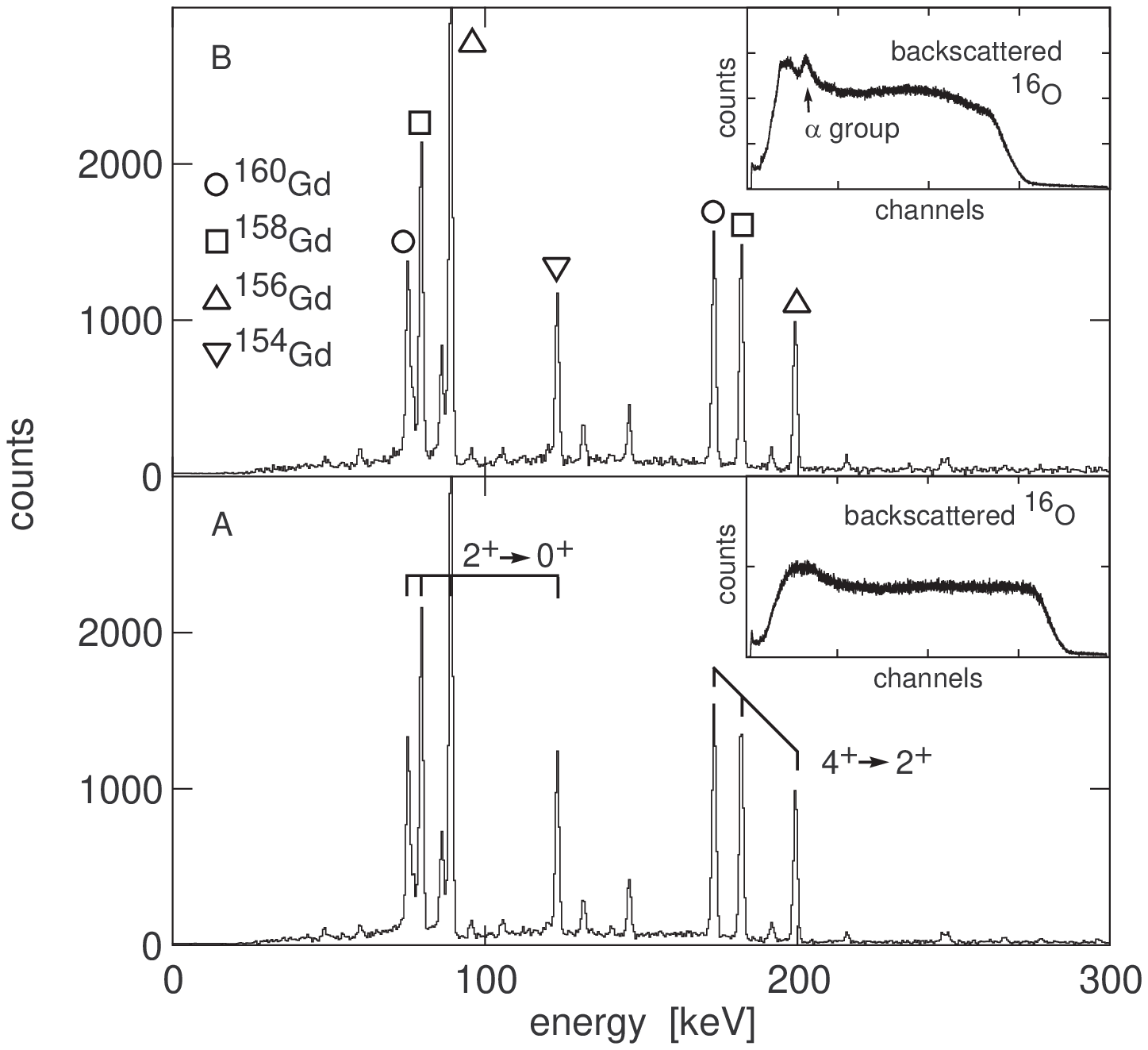}}
\caption{$\gamma$-ray spectra recorded in coincidence with
backscattered beam ions (see insets) for the two gadolinium targets.
Target B has a thin layer of Carbon which gives rise to the
$\alpha$-particle group. See Table~\protect \ref{tab:Egam} for
detailed identification of lines between 70 and 250~keV.}
\label{fig:gamspec}
\end{figure}

Particle and $\gamma$-ray spectra for the two targets are shown in
Fig~\ref{fig:gamspec}. The $\alpha$ particle group(s) at low
energies, appears in the spectrum for target B due to reactions on
the Carbon layer. The corresponding $\gamma$-ray spectra, measured
in coincidence with the detected particles, are essentially
identical in the region of interest, i.e. below 250~keV.

Table~\ref{tab:Egam} lists the observed $\gamma$-ray transitions
having energies between 70 keV and  250~keV. Relative $\gamma$-ray
intensities at $\theta_\gamma = 65^\circ$ are also given. The
uncertainty in the relative intensities is better than 5\% of the
quoted value for the strongest lines and $\sim 20\%$ for the
weakest. The $2^+_1 \rightarrow 0^+_1$ transitions in the even
isotopes are sufficiently resolved from the close-by $7/2^-_1
\rightarrow 5/2^-_1$ transitions in $^{155}$Gd and $^{157}$Gd.

\begin{longtable}{cccr}
\caption{$\gamma$-ray lines from Coulomb excitation of $^{\rm
nat}$Gd with 40~MeV $^{16}$O.} \\
\hline \hline \multicolumn{1}{c}{$E_\gamma$ (keV)~$^a$} &
\multicolumn{1}{c}{Nucleus} & \multicolumn{1}{c}{$I^\pi_i
\rightarrow I^\pi_f$} &
\multicolumn{1}{c}{Intensity~$^b$}\\
 \hline \endfirsthead
\caption{(Continued)} \\
\hline \hline \multicolumn{1}{c}{$E_\gamma$ (keV)~$^a$} &
\multicolumn{1}{c}{Nucleus} & \multicolumn{1}{c}{$I^\pi_i
\rightarrow I^\pi_f$} &
\multicolumn{1}{c}{Intensity~$^b$}\\
 \hline \endhead
%
%
    75.3 &  $^{160}$Gd & $2^+ \rightarrow 0^+$     & 57.9 \\
    76.9 &  $^{157}$Gd & $7/2^- \rightarrow 5/2^-$     & 17.6 \\
    79.5 &  $^{158}$Gd & $2^+ \rightarrow 0^+$     & 80.4 \\
    86.6 &  $^{155}$Gd & $7/2^- \rightarrow 5/2^-$     & 25.0 \\
    89.0 &  $^{156}$Gd & $2^+ \rightarrow 0^+$     & 100  \\
    95.7 &  $^{157}$Gd & $9/2^- \rightarrow 7/2^-$     &  2.8 \\
   105.3 &  $^{155}$Gd & $9/2^- \rightarrow 7/2^-$     &  2.6 \\
   120.1 &  $^{157}$Gd & $11/2^- \rightarrow 9/2^-$     &  1.8 \\
   123.1 &  $^{154}$Gd & $2^+ \rightarrow 0^+$     & 22.0 \\
   131.4 &  $^{157}$Gd & $7/2^- \rightarrow 3/2^-$     &  5.3 \\
   140.6 &  $^{155}$Gd & $11/2^- \rightarrow 9/2^-$     &  0.9 \\
   146.1 &  $^{155}$Gd & $7/2^- \rightarrow 3/2^-$     &  7.4 \\
   172.8 &  $^{157}$Gd & $9/2^- \rightarrow 5/2^-$     &  2.6~$^c$\\
   173.2 &  $^{160}$Gd & $4^+ \rightarrow 2^+$     & 33.0 \\
   181.9 &  $^{158}$Gd & $4^+ \rightarrow 2^+$     & 31.4 \\
   191.7 &  $^{155}$Gd & $9/2^- \rightarrow 5/2^-$     &  2.5 \\
   199.2 &  $^{156}$Gd & $4^+ \rightarrow 2^+$     & 22.8 \\
   215.6 &  $^{157}$Gd & $11/2^- \rightarrow 7/2^-$     &  1.4 \\
   246.2 &  $^{155}$Gd & $11/2^- \rightarrow 7/2^-$     &  1.6 \\
   247.9 &  $^{154}$Gd & $4^+ \rightarrow 2^+$     &  1.6 \\ \hline
   \hline
%
\multicolumn{4}{l}{$^a$~{$\gamma$-ray energies from Refs.~\protect
\cite{nds154,nds155,nds156,nds157,nds158,nds160}.}}\\
\multicolumn{4}{l}{$^b$~Relative $\gamma$-ray intensity at
$\theta_\gamma = 65^\circ$.}\\
\multicolumn{4}{l}{$^c$~{This transition was not directly observed;
see text.}}\\
 \label{tab:Egam}
\end{longtable}

The 172.8~keV, $9/2^- \rightarrow 5/2^-$ transition in $^{157}$Gd
was not observed, but its presence was inferred from the observation
of the $9/2^- \rightarrow 7/2^-$ transition and the known branching
ratio \cite{nds157}. A small correction to the precession data was
made to account for the effect of the overlap between the 173 keV
$4^+_1 \rightarrow 2^+_1$ transition in $^{160}$Gd and this much
weaker transition in $^{157}$Gd.

\subsection{Adopted lifetimes, quadrupole moments and $g$ factors}
\label{sect:adopted}

The level lifetimes, quadrupole moments and $g$~factors required for
the following analysis are summarized in Tables \ref{tab:tau} and
\ref{tab:g-Q}. The present analysis assumes that, in each isotope,
$g(2_1^+) = g(4_1^+) = g(6_1^+)$. This assumption is justified both
by experimental evidence and theoretical expectations \cite{stu91}.

\begin{table}
\caption{Lifetimes in the Gd isotopes}
\begin{tabular}{cccccc} \hline \hline
\multicolumn{1}{c}{Nucleus} & \multicolumn{1}{c}{$Q_0$~$^a$ (b) } &
\multicolumn{1}{c}{level} & \multicolumn{3}{c}{mean life (ps)} \\
\cline{4-6}
 & & & rotor & experiment & Ref.\\ \hline
$^{154}$Gd & 6.25  & 2$^+$ & 1710 & 1710(20) & \protect \cite{raman}\\
$^{156}$Gd & 6.83  & 2$^+$ & 3270 & 3270(60) & \protect \cite{raman}\\
$^{156}$Gd & 6.83  & 4$^+$ & 160 &  161.5(2.5) & \protect \cite{nds156}\\
$^{158}$Gd & 7.10  & 2$^+$ & 3730 & 3730(70) & \protect \cite{raman}\\
$^{158}$Gd & 7.10  & 4$^+$ &  219 &  214(3) & \protect \cite{nds158}\\
$^{160}$Gd & 7.27  & 2$^+$ & 3910 & 3910(80) & \protect \cite{raman}\\
$^{160}$Gd & 7.27  & 4$^+$ &  258 &  -\\ \hline \hline
\multicolumn{6}{l}{$^a$~Intrinsic quadrupole moments $Q_0$ are from
Ref.~\protect \cite{raman}.}\\
\end{tabular}
 \label{tab:tau}
\end{table}

\begin{table}
\caption{Quadrupole moments and $g$~factors of 2$^+_1$ states in the
Gd isotopes}
\begin{tabular}{ccccc} \hline \hline
\multicolumn{1}{c}{Nucleus} & \multicolumn{1}{c}{$E_x$ (keV)} &
\multicolumn{2}{c}{$Q$ (b) } &
\multicolumn{1}{c}{$g$~$^a$} \\
\cline{3-4}
 & &  rotor~$^b$ & experiment~$^c$ \\ \hline
$^{154}$Gd & 75  & -1.79 & -1.82 (4) & 0.430 (30) \\
$^{156}$Gd & 79  & -1.95 & -1.93 (4) & 0.387 (4) \\
$^{158}$Gd & 89  & -2.03 & -2.01 (4) & 0.381 (4) \\
$^{160}$Gd & 123 & -2.08 & -2.08 (4) & 0.364 (17) \\ \hline \hline
\multicolumn{5}{l}{$^a$~Adopted $g$~factors from the tabulation in
Ref.~\protect \cite{stu95}.}\\
\multicolumn{5}{l}{$^b$~Rotor model spectroscopic quadrupole moments
derived from the $Q_0$ values in Table~\protect \ref{tab:tau}.}\\
\multicolumn{5}{l}{$^c$~Quadrupole moments are from muonic atom
 x-ray measurements \protect \cite{lau83}.}\\
\end{tabular}
 \label{tab:g-Q}
\end{table}

The necessary lifetimes and quadrupole moments have been measured
for all but the 4$^+_1$ state of $^{160}$Gd. Since the rotor model
gives an excellent description of the measured lifetimes and
quadrupole moments in $^{156}$Gd and $^{158}$Gd, and $^{160}$Gd is
even more deformed than these isotopes, the rotor lifetime is
adopted for the 4$^+_1$ state in $^{160}$Gd. Experimental values are
used in all other cases.

\section{Magnetization measurements} \label{sect:magmeas}

The magnetizations of the gadolinium foils were measured off line
with the Rutgers magnetometer \cite{piq89}. For these measurements
samples were cut from the same rolled and annealed foils as those
used to make the targets. The results are summarized in
Table~\ref{tab:magnetom}. For these foils, and many similar foils
prepared by rolling and vacuum annealing, the magnetization is found
to vary with both temperature and applied field. Within a few
percent the magnetizations of the present foils track those of a
single crystal magnetized along the $b$ axis \cite{nigh}.

\begin{table}
\caption{Results of off-line magnetization measurements~$^a$.}
\begin{tabular}{cccccc} \hline \hline
\multicolumn{1}{c}{Sample} & \multicolumn{1}{c}{Temp} &
\multicolumn{1}{c}{$B_{\rm ext}$} & \multicolumn{1}{c}{$M$} &
\multicolumn{1}{c}{$\sigma$} &\multicolumn{1}{c}{$\sigma_{\rm sc}$}
\\
 &\multicolumn{1}{c}{(K)} & \multicolumn{1}{c}{(Oe)}
 &\multicolumn{1}{c}{(Gauss/cm$^3$)} & \multicolumn{1}{c}{(Gauss/g)}
 &\multicolumn{1}{c}{(Gauss/g)} \\
 \hline
A & 100 & 500 & 1523 & 193 & $\sim 205$ \\
B   &  50 & 783 & 1400 & 177 & $\sim 190$ \\
B   &  77 & 633 & 1618 & 205 & $\sim 200$ \\ \hline \hline
\\
\multicolumn{6}{l}{\parbox{4in}{$^a$~$M$ is the magnetic moment per
cm$^3$ while $\sigma$ is the magnetic moment per gram, thus $M =
\sigma \rho$, where $\rho$ is the density in g/cm$^3$. $\sigma_{\rm
sc}$ is the magnetization of a single crystal magnetized along the
$b$ axis as read off Fig.~5 in Ref.~\protect \cite{nigh}. The
quantities are given in cgs units for ease of comparison with
Ref.~\protect \cite{nigh}. The uncertainties in the measured
magnetizations are of the order of 3\%. There is an uncertainty of
the order of 5\% associated with reading $\sigma_{\rm sc}$ from the
figure. }}\\ &\\
\end{tabular}
 \label{tab:magnetom}
\end{table}

\begin{figure}
\begin{center}
    \resizebox{0.45\textwidth}{!}{
  \includegraphics[height=.8\textheight]{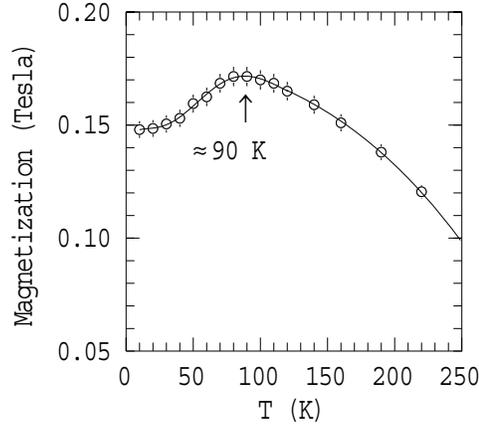}}
\caption{Measured magnetization as a function of temperature for a
5.3 mg/cm$^2$ gadolinium foil \protect \cite{rob99}. The line is
drawn to guide the eye and the error bars represent the uncertainty
in the absolute magnetization. The external polarizing field was
0.09~T. This magnetization curve is very similar to that of a single
crystal magnetized along the $b$ axis \protect \cite{nigh}. The
arrow indicates the approximate temperature at which many in-beam
measurements are performed.} \label{fig:magplot}
\end{center}
\end{figure}

Rolled and annealed thin Gd foils, as used in nuclear experiments,
typically have a texture such that the foil resembles a quasi-single
crystal with the $c$ axis perpendicular to the plane of the foil
(i.e. the basal planes of the microcrystals are in the plane of the
foil) \cite{bar66}. This texture is confirmed by x-ray diffraction
measurements \cite{ADFAxray}, and by the fact that the magnetization
curve versus temperature resembles that of a single crystal
magnetized along the $b$~axis. Fig.~\ref{fig:magplot} shows the
results of a detailed study of the magnetization of a 5.3 mg/cm$^2$
Gd foil as a function of temperature, which is very similar to that
of a single crystal magnetized along the $b$ axis \cite{nigh}.

The variation of the single-crystal magnetization with the external
field is shown in Fig.~5 of Ref.~\cite{nigh}. In the region around
90 K a factor of two change in the external field (from 0.05 to 0.1
T) changes the magnetization by only 5\%. The primary focus of this
paper is therefore on the temperature variation of the
magnetization.

\section{Unperturbed angular correlations} \label{sect:AC}

\subsection{Formalism} \label{sect:ACformalism}

\begin{figure} \begin{center}
    \resizebox{1.0\textwidth}{!}{
  \includegraphics[height=.8\textheight]{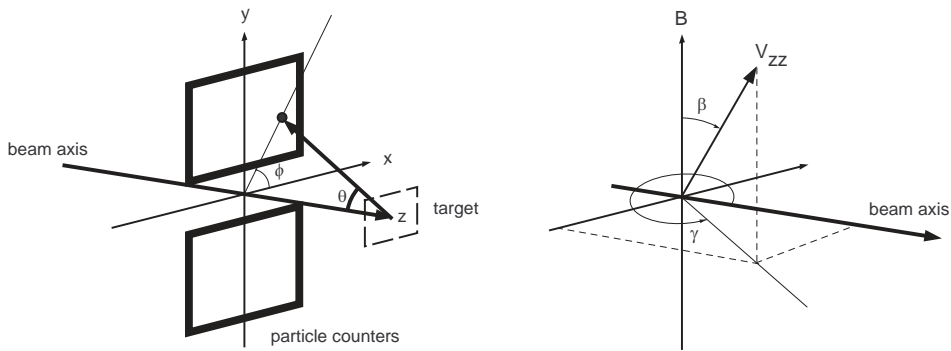}}
\caption{{\em Left:} Schematic of particle detector defining the
co-ordinate frame. The beam is along the $z$-axis and the magnetic
field direction is along the $y$-axis. The $\gamma$-ray detectors
are in the $xz$-plane. {\em Right:} Definitions of the spherical
polar angles $(\beta, \gamma)$ which specify the direction of the
electric field gradient (EFG), $\bm V_{zz}$, with respect to the
magnetic field direction and the beam axis. The EFG in the
gadolinium target foils studied here is aligned predominantly along
the beam axis, i.e. $\beta = 90^\circ$, $\gamma = 0^\circ$.}
\label{fig:sketch}
\end{center} \end{figure}

The theoretical expression for the unperturbed angular correlation
after Coulomb excitation can be written as (see Ref.~\cite{stu02,AW}
and references therein)
\begin{equation}
W(\theta_\gamma, \phi_\gamma) = \sum_{k,q} \sqrt{(2k+1)} \langle
\rho_{kq}(\theta,\phi) \rangle F_k Q_k
D^{k*}_{q0}(\phi_\gamma,\theta_\gamma,0) , \label{eq:wtheta}
\end{equation}
where $\langle \rho_{kq}(\theta,\phi) \rangle$ is the statistical
tensor, which defines the spin alignment of the initial state, and
which depends on the particle scattering angles $(\theta,\phi)$ and
the geometry of the particle detector. $F_k$ represents the usual
$F$-coefficients for the $\gamma$-ray transition, $Q_k$ is the
attenuation factor for the finite size of the $\gamma$-ray detector,
and $D^k_{q0}(\phi_\gamma,\theta_\gamma,0)$ is the rotation matrix,
which depends on the $\gamma$-ray detection angles
$(\theta_\gamma,\phi_\gamma)$. In the applications of interest $k
=0,2,4$. The co-ordinate frame is right-handed with the beam along
the positive $z$-axis as shown in Fig.~\ref{fig:sketch}. In the
present work the $\gamma$-ray detectors are in the $xz$ plane, thus
$\phi_\gamma = 0$, and the rotation matrix is equivalent to an
associated Legendre polynomial. The statistical tensors, and hence
the unperturbed angular correlations, can be calculated accurately
for the reaction geometry using the theory of Coulomb excitation.


The Coulomb excitation calculations performed here are based on the
de Boer-Winther code \cite{wdb}. In this code the statistical
tensors are evaluated in the particle-scattering plane. To calculate
the angular correlations requires the tensors corresponding to
scattering at angle $\phi$ as defined in Fig.~\ref{fig:sketch}.
These are given by
\begin{eqnarray}
\rho_{kq}(\theta, \phi) &=& \sum_{q^\prime} \rho_{k
q^\prime}(\theta, 0) D^k_{q^\prime q}(\phi,0,0) \\
 &=& \rho_{k q}(\theta, 0)
 {\rm e}^{i q\phi},
\end{eqnarray}
where $\rho_{k q}(\theta, 0)$ are from the de Boer-Winther
calculation. Thus the required average statistical tensor at a given
beam energy is given by
\begin{equation}
\langle \rho_{kq} \rangle = \frac{\int_{\theta} \int_{\phi} \rho_{k
q}(\theta, 0) {\rm e}^{i q\phi} \frac{{\rm d} \sigma}{{\rm d}
\Omega} {\rm d} \Omega }
 {\int_{\theta} \int_{\phi}
 \frac{{\rm d} \sigma}{{\rm d} \Omega} {\rm d} \Omega },
\end{equation}
where the integrals are over the dimensions of the particle detector
and $\frac{{\rm d} \sigma}{{\rm d} \Omega}$ is the cross section for
Coulomb excitation corresponding to the scattering angle $\theta$.
In the geometry used here (Fig.~\ref{fig:sketch}) there are two
particle detectors placed symmetrically about the beam axis such
that the numerical integration can be limited to the positive
quadrant, $0^\circ \leq \phi \leq 90^\circ$. The factor $ {\rm
e}^{iq \phi}$ can then be replaced by $( {\rm e}^{iq \phi} + {\rm
e}^{-iq \phi} + {\rm e}^{iq (\phi+\pi)} + {\rm e}^{-iq (\phi+\pi)}
)/4$, which is $\cos q \phi$ if $q$ is even and is zero if $q$ is
odd.

To obtain the statistical tensors of direct relevance to the present
experiments, a further integration was performed to average over the
energy loss of the beam in the target. A correction for feeding from
higher states in the ground-state band was also made. Since the
feeding path is only along the ground-state band, the statistical
tensor of the fed state $i$ can be evaluated iteratively using
\begin{equation}
\rho^{\rm fed}_{kq}(i) = \frac{\rho_{kq}(i) P^{\rm direct}(i) +
U_k(i+1 \rightarrow i) \rho^{\rm fed}_{kq}(i+1) P^{\rm
total}(i+1)}{P^{\rm direct}(i) + P^{\rm total}(i+1)},
\end{equation}
where $\rho_{kq}(i)$ is the unfed statistical tensor for the state
and $P^{\rm direct}(i)$ is the direct population of the state by
Coulomb excitation. $P^{\rm total}(i+1)$ is the total population of
the ground-band level above level $i$, including direct excitation
and feeding contributions, if any. $U_k(i+1 \rightarrow i)$ is the
$U$-coefficient for the $i+1 \rightarrow i$ transition \cite{yam67}.
These feeding corrections are small in the present work. In all
cases feeding from states with $I > 2$ in the ground-state band
contributes less than 7\% of the total intensity in the $2^+_1
\rightarrow 0^+_1$ transition.

The resultant nonzero statistical tensors for the 2$^+_1$ state of
$^{156}$Gd are shown in Table~\ref{tab:tensors}. For an annular
counter only the $q=0$ tensors are non zero. The broken azimuthal
symmetry gives rise to the finite $\rho_{k q}$ values for $q \neq
0$, however these terms are small in the present case because the
scattering angle remains near 180$^\circ$ and the spin of the
excited state is aligned predominantly in the plane perpendicular to
the beam.

\begin{table}
\caption{Statistical tensors for the 2$^+_1$ state of $^{156}$Gd.}
\begin{tabular}{ccc} \hline \hline
        $k$  &  $q$   & $\rho_{k,q}$\\ \hline
        0 &   0  &    1.0000\\
        2 &   2  &    0.0061\\
        2 &   0  &   -0.5033\\
        4 &   4  &    0.0002\\
        4 &   2  &   -0.0304\\
        4 &   0  &    0.4404\\ \hline \hline
\end{tabular}
\label{tab:tensors}
\end{table}

\subsection{Angular correlation results and discussion}
\label{sect:ACresults}

\begin{figure}  \begin{center}
    \resizebox{0.45\textwidth}{!}{
  \includegraphics[height=.8\textheight]{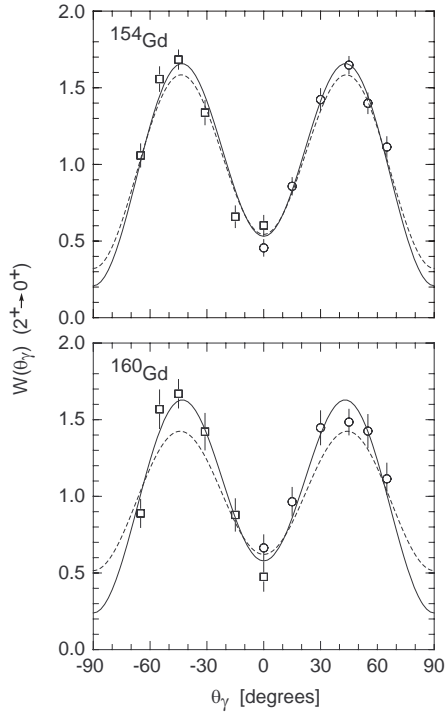}}
\caption{Unperturbed angular correlations for the $2^+_1 \rightarrow
0^+_1$ transitions in $^{154}$Gd and $^{156}$Gd. Circles (squares)
represent data taken with the detector in the positive (negative)
quadrant. The solid line is the theoretical angular correlation
given by Eq.~(\protect \ref{eq:wtheta}). The dotted line shows the
effect of an isotropic electric-field gradient ($V_{zz} = 3.44
\times 10^{17}$~V/cm$^2$). The good agreement between the data and
the solid lines implies that the electric-field gradient is
predominantly directed along the beam direction, and hence that the
gadolinium foil is a quasi single crystal with the $c$~axis
perpendicular to the plane of the foil.} \label{fig:AD2plus}
 \end{center} \end{figure}

\begin{figure}  \begin{center}
    \resizebox{0.45\textwidth}{!}{
  \includegraphics[height=.8\textheight]{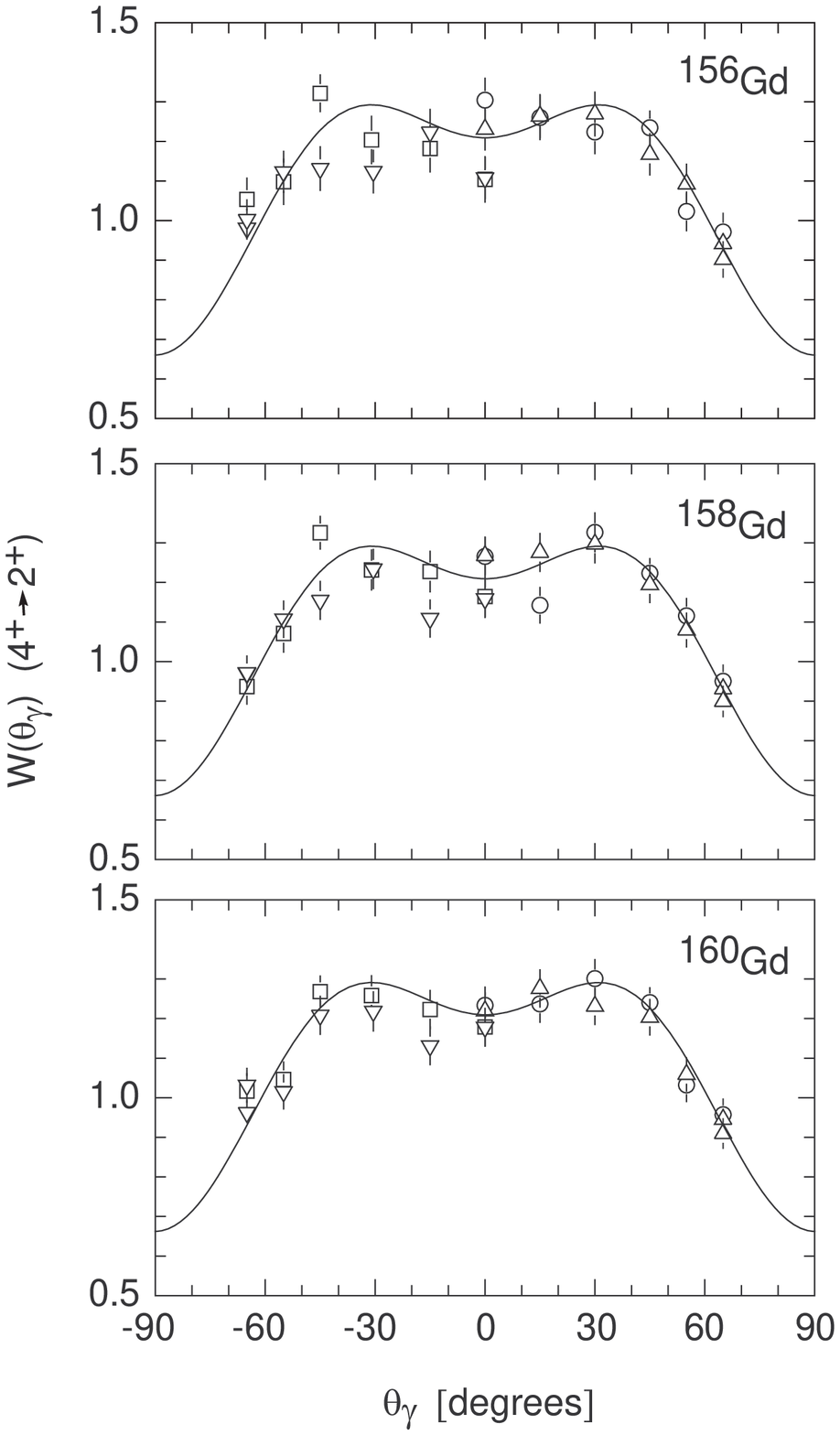}}
\caption{Unperturbed angular correlations for the  $4^+_1
\rightarrow 2^+_1$ transitions in $^{156,158,160}$Gd. Circles and
squares represent data taken at room temperature in run II. The
triangles represent data taken at a nominal target temperature of
90~K in run III. As in Fig.~\protect \ref{fig:AD2plus} different
symbols are used to designate data taken with the detectors in the
positive and negative quadrants.} \label{fig:AD4plus}
 \end{center} \end{figure}

Figures \ref{fig:AD2plus} and \ref{fig:AD4plus} show examples of
comparisons between the calculated unperturbed angular correlations
and the data. The angular correlations for the $2^+_1 \rightarrow
0^+_1$ transitions shown in Fig.~\ref{fig:AD2plus} were obtained at
room temperature (run I), above the Curie temperature. Thus there is
no magnetic-dipole perturbation, but there is expected to be an
electric field gradient (EFG). In Fig.~\ref{fig:AD2plus} the dotted
lines indicate the angular correlation anticipated for a target with
the expected electric field gradient of $V_{zz} = 3.44 \times
10^{17}$ V/cm$^2$ distributed isotropically. There is no evidence
for any attenuation of the anisotropy due to electric field
gradients. This observation again confirms that these rolled and
annealed foils have a texture such that the basal planes are
perpendicular to the beam, in the plane of the foil. There is no
perturbation because the $c$ axis, and hence the EFG, is along the
beam direction, perpendicular to the plane of spin alignment.

The data shown for the $4^+_1 \rightarrow 2^+_1$ transitions in
Fig.~\ref{fig:AD4plus} include data obtained at both room
temperature and at 90~K (runs I and II), where magnetic
perturbations are present. The calculated and measured angular
correlations are in agreement. There is no observable difference
between the measurements at 90~K and 300~K because the unperturbed
angular correlations can be recovered by adding together the data
for both directions of the external field. This procedure cancels
out the relatively small rotation of the radiation pattern which
changes direction when the external field direction is reversed.
(The same procedure cannot be applied for the 2$^+_1$ states because
the perturbations are much too large.)

\section{Perturbed angular correlations} \label{sect:pac}

\subsection{Formalism} \label{sect:pacform}

Since the perturbation of the angular correlation stems from changes
in the distribution of the nuclear spins as specified by changes in
the statistical tensors, the expression for the perturbed angular
correlation has the same form as Eq.~(\ref{eq:wtheta}), with the
statistical tensors $\langle \rho_{k q} \rangle$ replaced by values
that correspond to the perturbed spin distribution. The effect of
the transient-field precession, which acts as a pure rotation around
the direction of the magnetic field, is applied first. As shown in
Fig.~\ref{fig:sketch}, the magnetic field is directed along the
$y$-axis. The statistical tensor therefore becomes
\begin{equation} \protect \label{eq:rhorotate}
 \rho_{k q} = \sum_{Q} \langle \rho_{k Q} \rangle
 D^k_{q Q}(0,\Delta \theta_{\rm tf}, 0),
\end{equation}
where $\langle \rho_{k Q} \rangle$ are the unperturbed tensors. The
sign of $\Delta \theta_{\rm tf}$ is reversed when the direction of
the polarizing field is reversed.

After the transient-field precession has been applied, a combined
electric-quadrupole and magnetic-dipole interaction, due to the
static fields in the gadolinium host matrix, is allowed to perturb
the statistical tensor. The perturbed tensors, $\rho_{k^\prime
q^\prime}$, are derived from the tensors, $\rho_{k q}$, using
\begin{equation} \label{eq:rhopert}
\rho_{k^\prime q^\prime} = \sum_{k,q} \rho_{k,q} [G^{ q q^\prime}_{k
k^\prime}]^* \sqrt{\frac{(2k +1)}{(2k^\prime + 1)}},
\end{equation}
where $[G^{q q^\prime}_{k k^\prime}]^*$ is the perturbation factor
for the combined magnetic and electric hyperfine interactions.
$[G^{q q^\prime}_{k k^\prime}]^*$ can be related to the perturbation
factors denoted ${\rm III}^{q q^\prime}_{k k^\prime}(\beta,\gamma)$,
given by Alder {\em et al.} \cite{ald63,mat62} for combined electric
and magnetic interactions:
\begin{equation}
[G^{ q q^\prime}_{k k^\prime}(t)]^* = \frac{\sqrt{(2k+1) (2k^\prime
+1)}}{(2I+1)} \sum_{Q,Q^\prime} D^{k * }_{q Q}(\frac{\pi}{2},
\frac{\pi}{2}, \pi) {\rm III}^{Q Q^\prime}_{k
k^\prime}(\beta,\gamma) D^{k^\prime }_{q^\prime
Q^\prime}(\frac{\pi}{2}, \frac{\pi}{2}, \pi).
\end{equation}
The Euler (or spherical polar) angles $(\beta, \gamma)$ specify the
orientation of the electric field gradient with respect to the
magnetic field direction, which is along the $y$ axis in
Fig.~\ref{fig:sketch}. $\beta$ is the polar angle between the
directions of the magnetic field and the electric-field gradient.
The azimuthal angle $\gamma$ is measured from the beam axis in the
horizontal plane, i.e. the $xz$-plane in Fig.~\ref{fig:sketch}.
Although it is not usually written explicitly, ${\rm III}^{Q
Q^\prime}_{k k^\prime}(\beta,\gamma)$ is a function of $\omega \tau$
and $\omega_Q \tau$, where $\omega \tau$ is the magnetic dipole
precession angle as defined above and $\omega_Q \tau$ is the
electric quadrupole precession, where $\omega_Q$ is
\begin{equation}
\omega_Q = \frac{eQV_{zz}}{4 I (2I-1) \hbar}.
 \protect \label{eq:omQ}
\end{equation}

\subsection{Data analysis procedures}\label{sect:analysis}

As a first approximation, the experimental total precession angles
for the 4$^+_1$ states, $\Delta \Theta$, can be extracted from the
field-up/field-down data by conventional procedures
\cite{ben80,spe02} in which the experimental precession angle is
related to the field up/down counting asymmetry $\epsilon$ by the
expression
\begin{equation}
\Delta \Theta = {\epsilon} / {S} , \label{eq:dtheta}
\end{equation}
where $S=(1/W)(dW/d\theta)$ is the logarithmic derivative of the
angular correlation at the detection angle $+\theta_\gamma$ and
\begin{equation}
\epsilon = (1 - \rho)  / (1 + \rho ) . \protect\label{eq:eps}
\end{equation}
The `double ratio' $\rho$ is derived from the counting rates in the
detectors at $\pm \theta _{\gamma}$, $N(\pm \theta _{\gamma})$, for
field up ($\uparrow$) and down ($\downarrow$) by
\begin{equation}
\rho = \sqrt{ \frac{N(+\theta _{\gamma},\uparrow) } {N(+\theta
_{\gamma}, \downarrow)} \frac{N(-\theta _{\gamma}, \downarrow)}
{N(-\theta _{\gamma}, \uparrow)}  }. \label{eq:rho}
\end{equation}
Note that the factors due to integrated beam current, cross sections
and detector efficiencies cancel out so that
\begin{equation}
\rho = W(+\theta _{\gamma},\uparrow) / W(+\theta _{\gamma},
\downarrow)
\end{equation}
and hence $\epsilon$ is formally equivalent to
\begin{equation}
\epsilon(\theta _{\gamma}) = \frac{W(+\theta _{\gamma}, \downarrow)
- W(+\theta _{\gamma},\uparrow)}{W(+\theta _{\gamma}, \downarrow) +
W(+\theta _{\gamma},\uparrow)}. \label{eq:epsilon}
\end{equation}

For the longer-lived 2$^+_1$ states, where the perturbations are
much larger, this procedure does not apply. Indeed, it also
underestimates the precessions of the 4$^+_1$ states by up to $\sim
10\%$. The procedure used here to analyze both the 2$^+_1$ and
4$^+_1$ data is to begin by forming the experimental double ratio,
$\rho$, and asymmetry parameter $\epsilon$ as usual, so that the
beam current, cross section and efficiency factors cancel. The
experimental value of $\omega \tau$ is then extracted by fitting the
experimental values of $\epsilon$ (Eq.~(\ref{eq:eps}) and
Eq.~(\ref{eq:rho})) to theoretical values of $\epsilon$ evaluated
using Eq.~(\ref{eq:epsilon}).

\subsection{Electric quadrupole and magnetic dipole
perturbations} \label{sect:E-B}

\begin{figure}  \begin{center}
    \resizebox{0.8\textwidth}{!}{
  \includegraphics[height=.8\textheight]{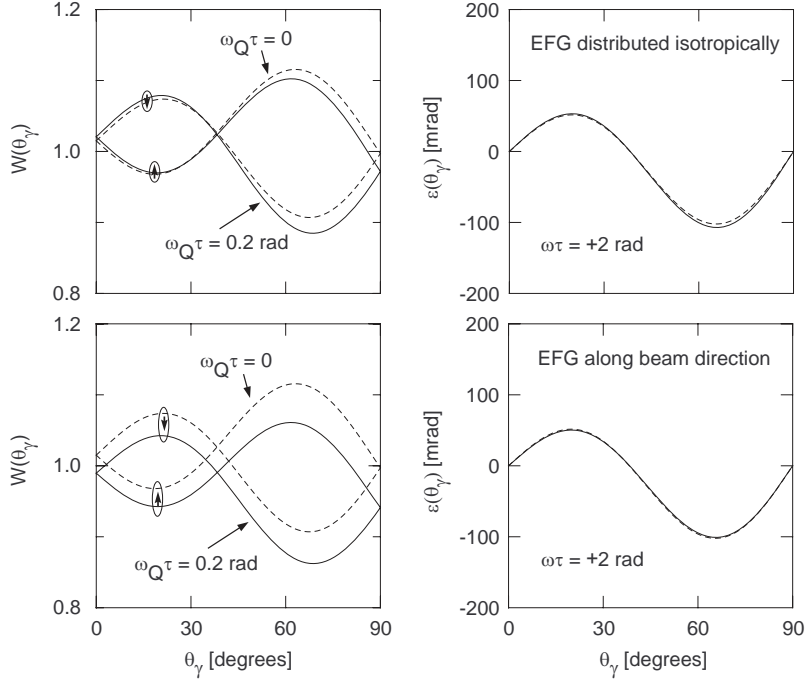}}
\caption{Perturbed angular correlations for a hypothetical $2^+_1
\rightarrow 0^+_1$ transition which experiences a combined
magnetic-dipole and electric-quadrupole interaction. The left-hand
panels show the angular correlation, $W(\theta_\gamma)$, while the
right-hand panels show the asymmetry, $\epsilon(\theta_\gamma)$,
derived from double ratios as described in the text concerning
Eq.(\protect \ref{eq:eps}) and Eq.(\protect \ref{eq:rho}). In all
panels the magnetic precession is $\omega \tau = +2$~rad for `field
up'. The field directions are indicated by the arrows. Solid lines
show the case where an electric-field gradient, which causes a
quadrupole precession of $\omega_Q \tau = 0.2$~rad, is also present.
For reference the dotted lines show the $\omega_Q \tau = 0$ case. In
the upper two panels the electric-field gradient is distributed
isotropically, while in the lower panels it is directed along the
beam axis, as is the case in the present measurements.}
\label{fig:wthetEFG}
 \end{center} \end{figure}

\begin{figure}  \begin{center}
    \resizebox{0.8\textwidth}{!}{
  \includegraphics[height=.8\textheight]{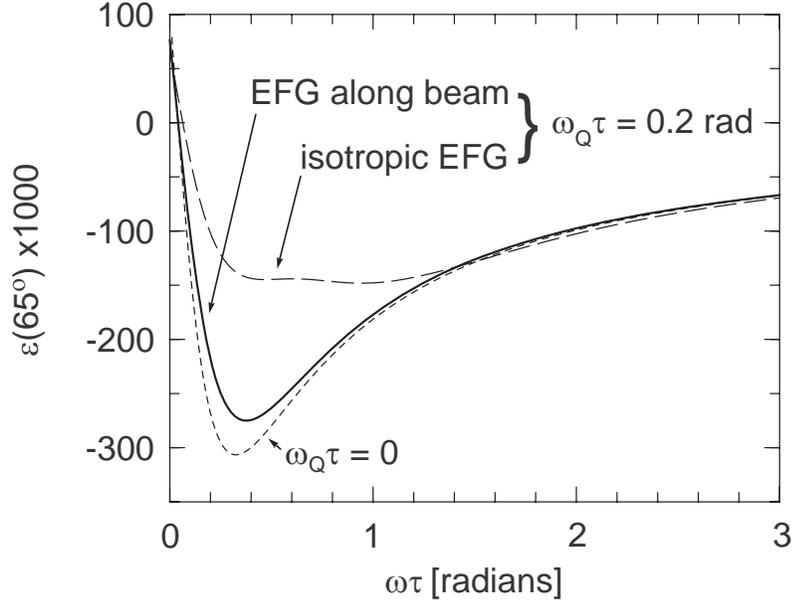}}
\caption{Asymmetry, $\epsilon(65^\circ)$, for a typical $2^+
\rightarrow 0^+$ transition, as a function of the magnetic
precession angle $\omega \tau$ when an electric-quadrupole
precession of $\omega_Q \tau = 0.2$~rad is present, either
distributed isotropically or directed along the beam direction (as
is the case in the present work). The case of $\omega_Q \tau = 0$ is
shown for reference. In the present work the $\omega \tau$ values
lie between $\sim 0.8$ and $\sim 2.2$ rad, with $\omega_Q \tau =
0.2$ being applicable when $\omega \tau \sim 2.5$ rad. See also
Fig.~\protect \ref{fig:epsOmT1}.} \label{fig:epsOmT}
 \end{center} \end{figure}

\begin{figure}  \begin{center}
    \resizebox{0.8\textwidth}{!}{
  \includegraphics[height=.8\textheight]{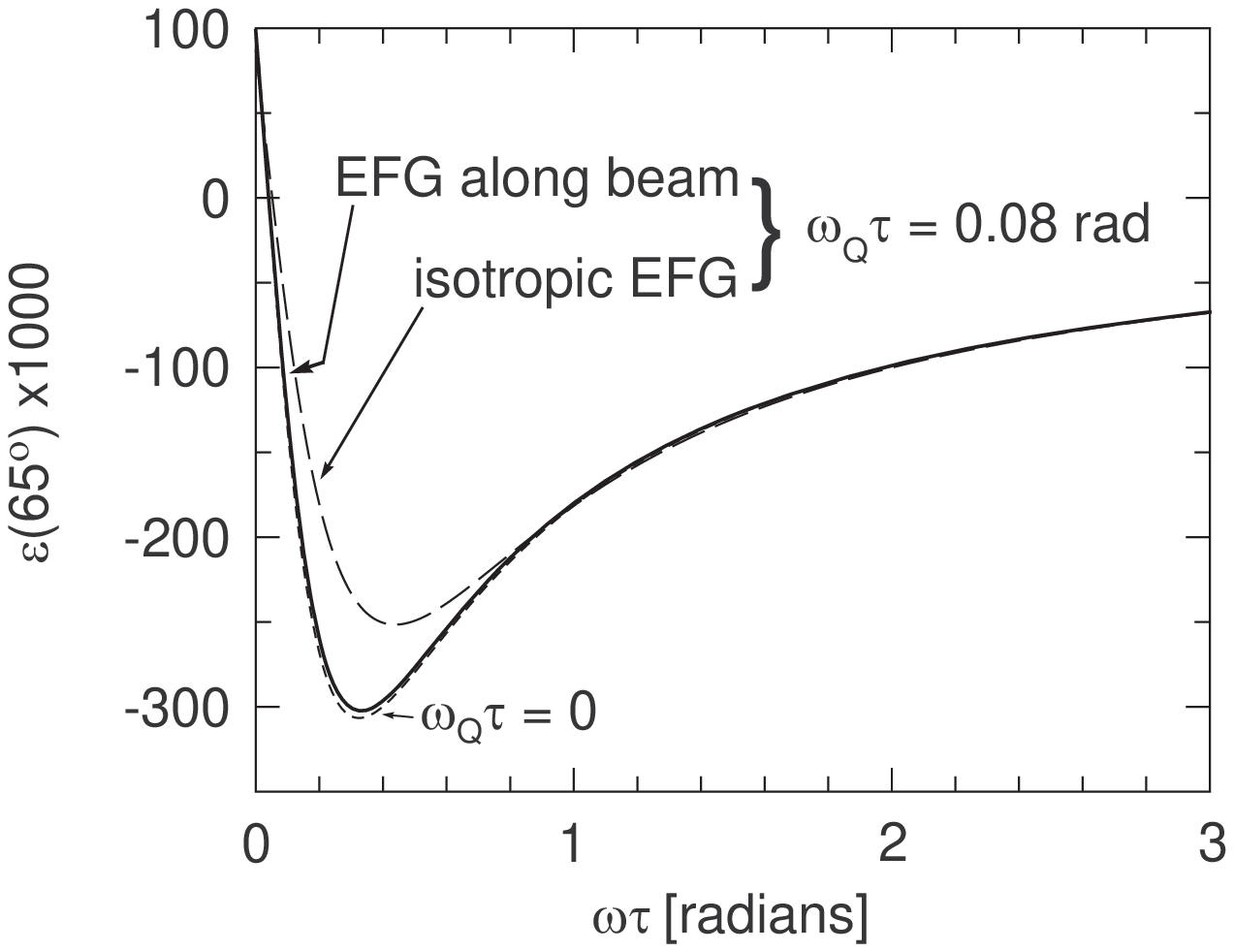}}
\caption{As for Fig.~\protect \ref{fig:epsOmT}, but for $\omega_Q
\tau = 0.08$~rad. In the present work $\omega_Q \tau = 0.08$~rad is
associated with $\omega \tau \sim 0.8$ rad. } \label{fig:epsOmT1}
 \end{center} \end{figure}

The combined effect of electric quadrupole and magnetic dipole
interactions on the angular correlation and the asymmetry $\epsilon$
is explored in Figs.~\ref{fig:wthetEFG}-\ref{fig:epsOmT1}.

In Fig.~\ref{fig:wthetEFG}, the transient-field precession, which is
relatively small, was set to zero, the magnetic dipole precession
angle was $\omega \tau = 2$~rad, and the electric quadrupole
precession was $\omega_Q \tau = 0.2$~rad. These values are near
those for $^{160}$Gd in the data presented below. Two extreme cases
are shown. In the upper panels of Fig.~\ref{fig:wthetEFG} the EFG is
assumed to be distributed isotropically, while in the lower panels
it is directed along the beam axis. As discussed above, the latter
case is very close to the real situation. Although the electric
field gradient is directed along the beam axis, perpendicular to the
initial spin orientation, such that it cannot perturb the angular
correlation on its own, it can have an observable effect on the
perturbed angular correlation when the magnetic dipole interaction
is also present, because the magnetic interaction moves the nuclear
spin out of the plane perpendicular to the beam. Thus the perturbed
angular correlations shown in the left panels of
Fig.~\ref{fig:wthetEFG} differ, depending on the direction and
magnitude of the electric field gradient.

Figures \ref{fig:epsOmT} and \ref{fig:epsOmT1} show the dependence
of the asymmetry $\epsilon (65^\circ)$ on the magnitude of the
magnetic perturbation $\omega \tau$ for a given strength of the
quadrupole interaction. In these figures a realistic value of the
transient-field precession was included, $\Delta \theta = -40$~mrad,
where the negative sign applies for `field up'. The value $\omega_Q
\tau = 0.2$ rad was chosen because it is slightly larger than the
value applicable for $^{160}$Gd, the largest considered here;
$\omega_Q \tau = 0.08$ rad is slightly larger than the value for
$^{154}$Gd. The effect of the electric-field gradient is apparent in
the region up to $\omega \tau \approx 1$ rad when the EFG is
isotropically distributed. However when $\omega \tau >> \omega_Q
\tau$, the asymmetries, $\epsilon(65^\circ)$, are hardly affected,
especially when the EFG is directed along the beam direction.

In the following analysis $\omega \tau$ values are extracted
assuming the EFG is directed along the beam direction, as was shown
to apply for our foils in Sections \ref{sect:magmeas} and
\ref{sect:AC}. $\omega_Q$ was evaluated assuming the electric-field
gradient from Ref.~\cite{bau75} and the experimental quadrupole
moments in Table \ref{tab:g-Q}. The parameters of the EFG are not
critical, however. In all cases considered here essentially the same
$\omega \tau$ values would be extracted if EFG effects were assumed
isotropic or even ignored altogether.

\subsection{Nuclei on damaged sites} \label{sect:damage}

The analysis procedures described above implicitly assume that all
of the implanted nuclei experience the same hyperfine magnetic field
and the same electric-field gradient. It is well known, however,
that after implantation into metals typically 5 - 10\% of the
implanted nuclei reside on damaged sites and therefore do not
experience the same hyperfine fields as those on substitutional
sites.

It will become apparent in the following that an analysis of the
present data assuming a unique implantation site leads to a
contradiction between the effective fields experienced by the 4$^+$
and 2$^+$ states. A two-site analysis is necessary to resolve this
contradiction.

For the analysis of integral perturbed angular correlations it is
usually sufficient to assume a two-site model in which the damaged
sites have no static hyperfine magnetic field. This assumption will
be adopted here with the additional assumption that there is also no
net electric field gradient on the damaged sites. It will become
apparent in the following discussion that these assumptions are not
critical because the number of damaged sites is small.

In a two-site model with a fraction of nuclei $f$ on field-free
sites after implantation, the perturbed angular correlation,
$W(\Delta \theta_{\rm tf}, \omega \tau, \omega_Q \tau)$ is replaced
by
\begin{equation}
(1-f)W(\Delta \theta_{\rm tf}, \omega \tau, \omega_Q \tau) + f
W(\Delta \theta_{\rm tf}, 0, 0). \label{eq:fff}
\end{equation}
Note that the nuclei that end up on field-free sites still
experience the transient field as they slow in the ferromagnetic
medium, but they do not experience any hyperfine interactions after
they come to rest.

\begin{figure}  \begin{center}
    \resizebox{0.8\textwidth}{!}{
  \includegraphics[height=.8\textheight]{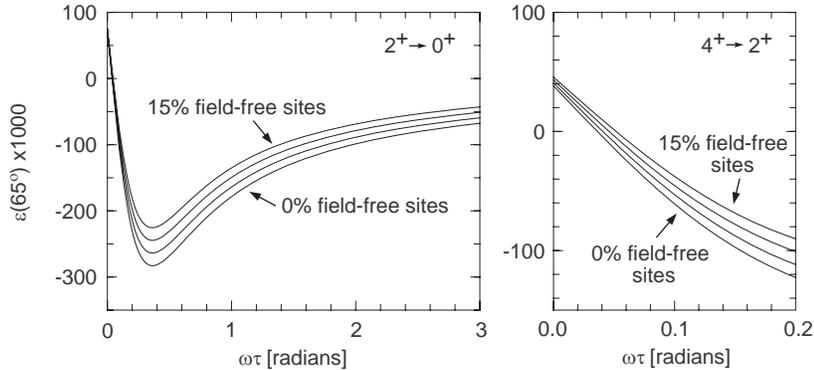}}
\caption{Asymmetry, $\epsilon(65^\circ)$, for typical $2^+
\rightarrow 0^+$ and $4^+ \rightarrow 2^+$ transitions, as a
function of the magnetic precession angle $\omega \tau$ when a
fraction of the implanted nuclei reside on field-free sites.
Calculations are shown for field-free fractions from 0\% to 15\% in
5\% steps.} \label{fig:epsblend}
 \end{center} \end{figure}

The effect of a fraction of damaged sites on the observed asymmetry
$\epsilon(65^\circ)$ is demonstrated in Fig.~\ref{fig:epsblend}. In
the present case a fraction of nuclei on field-free sites, which is
not taken into account, gives an apparently enhanced $B_{\rm st}$
for the 2$^+$ states and an apparently reduced $B_{\rm st}$ for the
4$^+$ states. The difference comes about because the longer-lived
2$^+$ states are in the region where an increase in $\omega \tau$
results in a decrease in the magnitude of $\epsilon$ whereas for the
4$^+$ states an increase in $\omega \tau$ results in an increase in
the magnitude of $\epsilon$.

\begin{landscape}
\begin{longtable}{ccccccccccc}
\caption{Summary of 2$^+$ precessions.} \\
\hline \hline
 \multicolumn{1}{c}{Run} &
 \multicolumn{1}{c}{$\Delta \theta_{\rm tf}$} &
 \multicolumn{1}{c}{$\omega_Q \tau$} &
 \multicolumn{1}{c}{$\epsilon(65^\circ)\times 10^3$}&
 \multicolumn{2}{c}{$\omega \tau(65^\circ)$} &
 \multicolumn{1}{c}{$\epsilon(120^\circ)\times 10^3$}&
 \multicolumn{2}{c}{$\omega \tau (120^\circ)$} &
 \multicolumn{2}{c}{$\langle \omega \tau \rangle$}\\
  & (mrad) & (mrad) & & \multicolumn{2}{c}{(mrad)} & &
\multicolumn{2}{c}{(mrad)}& \multicolumn{2}{c}{(mrad)}\\
%
  & & & & \multicolumn{1}{c}{1-site} &  \multicolumn{1}{c}{2-site}
  & & \multicolumn{1}{c}{1-site} &  \multicolumn{1}{c}{2-site}
  &  \multicolumn{1}{c}{1-site} & \multicolumn{1}{c}{2-site} \\
 \hline \endfirsthead
\caption{(Continued)} \\ \hline \hline
 \multicolumn{1}{c}{Run} &
 \multicolumn{1}{c}{$\Delta \theta_{\rm tf}$} &
 \multicolumn{1}{c}{$\omega_Q \tau$} &
 \multicolumn{1}{c}{$\epsilon(65^\circ)\times 10^3$}&
 \multicolumn{2}{c}{$\omega \tau(65^\circ)$} &
 \multicolumn{1}{c}{$\epsilon(120^\circ)\times 10^3$}&
 \multicolumn{2}{c}{$\omega \tau (120^\circ)$} &
 \multicolumn{2}{c}{$\langle \omega \tau \rangle$}\\
  & (mrad) & (mrad) & & \multicolumn{2}{c}{(mrad)} & &
\multicolumn{2}{c}{(mrad)}& \multicolumn{2}{c}{(mrad)}\\
  & & & & \multicolumn{1}{c}{1-site} &  \multicolumn{1}{c}{2-site}
  & & \multicolumn{1}{c}{1-site} &  \multicolumn{1}{c}{2-site}
   & \multicolumn{1}{c}{1-site} & \multicolumn{1}{c}{2-site} \\
 \hline \endhead
\multicolumn{11}{c}{$^{154}$Gd}\\
 II & $-35$ & 67 & $-204(14)$ &
904$^{+88}_{-78}$ & 698$^{+75}_{-69}$ &   221(16) &
784$^{+94}_{-86}$ &
537$^{+90}_{-103}$ & 849$^{+64}_{-58}$ & 640$^{+58}_{-57}$ \\
III& $-43$ & 67 & $-193(10)$ & 956$^{+68}_{-62}$ & 828$^{+61}_{-57}$
&   180(16) & 1030$^{+133}_{-115}$ &
 865$^{+117}_{-103}$ &  972$^{+61}_{-55}$ & 836$^{+54}_{-50}$ \\
IV & $-43$ & 67 & $-187(11)$ & 995$^{+80}_{-72}$ & 863$^{+71}_{-65}$
&   153(13) & 1242$^{+144}_{-123}$ & 1073$^{+123}_{-107}$ &
 1062$^{+70}_{-62}$ & 918$^{+61}_{-56}$\\ \hline
%
%
\multicolumn{11}{c}{$^{156}$Gd}\\
 II & $-31$ & 139 &
$-148(8)$ &1299$^{+87}_{-79}$ &1039$^{+71}_{-64}$
&   130(7) & 1529$^{+101}_{-91}$ &1168$^{+78}_{-70}$ &  1397$^{+66}_{-60}$ &1098$^{+53}_{-47}$ \\
III& $-39$ & 139 & $-120(5)$ &1635$^{+80}_{-74}$ &1435$^{+69}_{-64}$
&   117(7) & 1706$^{+125}_{-111}$ &1456$^{+105}_{-94}$ &  1657$^{+67}_{-62}$ &1441$^{+58}_{-53}$\\
IV & $-39$ & 139 & $-107(5)$ &1858$^{+100}_{-92}$
&1626$^{+86}_{-79}$ &   113(7) &1775$^{+133}_{-118}$
&1514$^{+112}_{-100}$&  1827$^{+80}_{-73}$ &1584$^{+68}_{-62}$\\
\hline
%
%
\multicolumn{11}{c}{$^{158}$Gd}\\
 II & $-31$ & 165 &
$-123(11)$ &1588$^{+176}_{-148}$ &1274$^{+140}_{-119}$ & 107(8) &
1893$^{+172}_{-148}$
&1447$^{+129}_{-112}$ &  1742$^{+123}_{-105}$ &1367$^{+95}_{-82}$ \\
III& $-38$ & 165 & $-100(6)$ &1984$^{+139}_{-123}$
&1740$^{+119}_{-106}$ & 94(9) & 2168$^{+254}_{-210}$
&1848$^{+210}_{-176}$ &  2029$^{+122}_{-106}$ &1767$^{+104}_{-91}$ \\
IV & $-38$ & 165 & $-91(8)$ & 2198$^{+229}_{-193}$ &
1922$^{+193}_{-164}$ &  101(8) &2001$^{+193}_{-165}$ &
1709$^{+161}_{-139}$ &  2084$^{+148}_{-125}$ & 1797$^{+124}_{-106}$
\\ \hline
&\\
%
\multicolumn{11}{c}{$^{160}$Gd}\\
 II & $-30$ & 177 &
$-111(12)$ & 1770$^{+238}_{-193}$ & 1424$^{+188}_{-154}$ &
   101(10) &2010$^{+246}_{-202}$ & 1541$^{+184}_{-153}$ &
    1885$^{+171}_{-140}$ & 1483$^{+131}_{-109}$ \\
III & $-36$ & 177 & $-87(7)$ & 2298$^{+216}_{-185}$ &
2016$^{+184}_{-158}$ &
   112(11) &1774$^{+220}_{-182}$ & 1524$^{+186}_{-154}$ &
    2036$^{+154}_{-130}$& 1768$^{+131}_{-110}$  \\
IV & $-36$ & 177 & $-73(8)$ & 2770$^{+360}_{-290}$ &
2415$^{+299}_{-244}$ &
   104(10) &1929$^{+231}_{-191}$ & 1656$^{+195}_{-162}$ &
    2179$^{+194}_{-160}$& 1885$^{+163}_{-135}$ \\ \hline \hline
\protect \label{tab:2+results}
\end{longtable}
\end{landscape}

\subsection{Results: 2$^+_1$ states} \label{sect:results2}

Table \ref{tab:2+results} summarizes the results of the present
measurements of the static-field precessions for the 2$^+_1$ states
in $^{154,156,158,160}$Gd. The data were analyzed adopting both a
single-site model (i.e. $f=0$ in Eq.~(\ref{eq:fff})) and a two site
model ($f \neq 0$). The procedure by which the field-free fraction
was determined for the two site-model analysis is described in
section \ref{sect:fff}.

\begin{figure} \begin{center}
    \resizebox{0.8\textwidth}{!}{
  \includegraphics[height=.8\textheight]{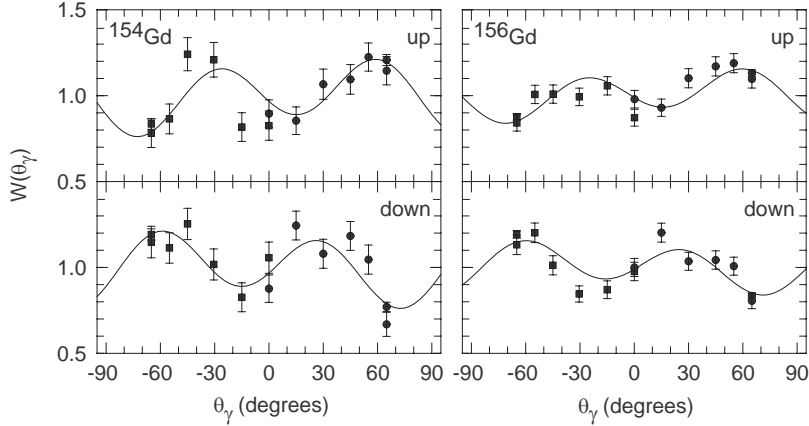}}
\caption{Perturbed angular correlations for the $2^+_1 \rightarrow
0^+_1$ transitions in $^{154}$Gd and $^{156}$Gd measured in run II.
The solid curves are calculated for the average single-site $\omega
\tau$ values in Table~\protect \ref{tab:2+results}. These data make
it clear that the $\omega \tau$ values for the 2$^+_1$ states are
much larger than 0.5 rad.} \label{fig:pacs}
\end{center} \end{figure}

While the $\omega \tau$ values Table~\ref{tab:2+results} were
extracted exclusively from the asymmetries, $\epsilon(65^\circ)$ and
$\epsilon(120^\circ)$, these values are consistent with the full
perturbed angular correlations, where measured. For example, the
perturbed angular correlation data for $^{154}$Gd and $^{156}$Gd
measured in run II are shown in Fig.~\ref{fig:pacs}, and compared
with the perturbed angular correlations corresponding to the $\omega
\tau$ values for the single-site model (see
Table~\ref{tab:2+results}).

To some extent the extracted magnetic-dipole precession angles,
$\omega \tau$, depend on the assumed transient-field precession
angle, $\Delta \theta_{\rm tf}$, and the quadrupole precession,
$\omega_Q \tau$. Figures~\ref{fig:wthetEFG} and \ref{fig:epsOmT1}
discussed in the previous subsection show that $\omega \tau$ is very
insensitive to reasonable assumed values of $\omega_Q \tau$. The
effect of the transient-field contribution is somewhat counter
intuitive at first sight: $\omega \tau$ decreases by $\sim 20$~mrad
when the magnitude of $\Delta \theta_{\rm tf}$ (which is negative)
increases by 10~mrad. (See Fig.~4 in Ref.~\cite{stu96} for a plot of
$\epsilon$ versus $\omega \tau$ showing the effect of the
transient-field contribution.) Because the $\omega \tau$ values are
so large, however, an increase in $\Delta \theta_{\rm tf}$ by $\sim
50$\% would lead to a change in $\omega \tau$ for $^{154}$Gd
($^{160}$Gd) by only $\sim 4$\% ($\sim 2$\%).

A fixed value of $\Delta \theta _{\rm tf} /g = -100$~mrad was
adopted for run IV, and scaled according to the foil magnetization
for runs II and III as obtained in a preliminary fit to the data.
This value was chosen because it is consistent with the previous
data \cite{ska72} once a correction is made for the difference in Gd
recoil velocities due to the difference in the beam energies. In
principle, $\Delta \theta_{\rm tf}$ could be obtained from the data
for the 4$^+$ states. Unfortunately, however, the present data for
the 4$^+$ states are not sufficiently precise to fit both the
static- and transient-field precessions as free parameters (see
below).

Results presented for the two-site model assume a field-free
fraction of 11.6\% in run II and 6.4\% in runs III and IV. These
values were determined by requiring consistency between the
extracted static-field strengths for the 2$^+_1$ and 4$^+_1$ states,
as described in section \ref{sect:fff}.

The effective static-field strengths derived from the average
precession angles given in Table~\ref{tab:2+results} are summarized
in Table~\ref{tab:2+fields}. As will be discussed below, the
differences in these effective static-field strengths from run to
run are attributed to changes in the magnetization of the gadolinium
foil target associated with different foil textures and different
beam heating effects.

\begin{landscape}
\begin{table*}
\caption{Effective static-field strengths derived from 2$^+$-state
precessions.}
\begin{tabular}{cccccccc} \hline \hline
\multicolumn{1}{c}{Nucleus} & \multicolumn{3}{c}{$B_{\rm st}$
(Tesla) Single-site analysis} & & \multicolumn{3}{c}{$B_{\rm st}$
(Tesla) Two-site analysis}
\\ \cline{2-4} \cline{6-8}
& Run II (A) & Run III (A) & Run IV (B) & & Run II (A) & Run III (A) & Run IV (B)\\
\hline $^{154}$Gd & $-24.1$(2.4) & $-27.6$(2.5) & $-30.2$(2.8) & &
$-18.2(2.1)$ & $-23.8(2.2)$ & $-26.1(2.5)$ \\
$^{156}$Gd & $-23.1$(1.1) & $-27.4$(1.2) & $-30.2$(1.4)& &
$-18.1(0.9)$ & $-23.8(1.0)$ & $-26.2(1.2)$ \\
$^{158}$Gd & $-25.6$(1.8) & $-29.8$(1.8) & $-30.6$(2.1) &
& $-20.1(1.4)$ & $-26.0(1.5)$ & $-26.4(1.8)$ \\
$^{160}$Gd & $-27.7$(2.7) & $-29.9$(2.6) & $-32.0$(3.1) &
& $-21.8(2.1)$ & $-26.0(2.2)$ & $-27.7(2.6)$ \\
 \cline{2-2} \cline{3-3} \cline{4-4} \cline{6-6} \cline{7-7} \cline{8-8}
average    & $-24.3$(0.8) & $-28.3$(0.9) & $-30.5$(1.0) & &
$-19.0$(0.7) & $-24.6$(0.8) & $-26.4$(0.9)
\\
\hline \hline
\end{tabular}
\protect\label{tab:2+fields}
\end{table*}
\end{landscape}

\subsection{Results: 4$^+_1$ states} \label{sect:results4}

Table \ref{tab:4+results} summarizes the results of the present
measurements on the 4$^+_1$ states in $^{156,158,160}$Gd. The 4$^+$
state in $^{154}$Gd was populated too weakly in the present work to
be analyzed. The precession angles $\omega \tau$ were obtained by
the rigorous procedure described in section \ref{sect:analysis}
assuming the same $\Delta \theta_{\rm tf}$ values as adopted for the
analysis of the 2$^+_1$ states (Table~\ref{tab:2+results}).

\begin{table*}
\caption{Summary of 4$^+_1$ state precessions.}
\begin{tabular}{cccccccc}
\hline \hline
 \multicolumn{1}{c}{Run} &
 \multicolumn{1}{c}{Target} &
 \multicolumn{1}{c}{$\Delta \theta_{\rm tf}$} &
 \multicolumn{1}{c}{$\epsilon(65^\circ)$} &
 \multicolumn{1}{c}{$\epsilon(120^\circ)$}&
 \multicolumn{2}{c}{$\langle \omega \tau \rangle$} &
 \multicolumn{1}{c}{$\langle \Delta \Theta \rangle/g$} \\
 &
  &
  &
 \multicolumn{1}{c}{$(\times 10^3$)} &
 \multicolumn{1}{c}{$(\times 10^3$)} &
 \multicolumn{2}{c}{(mrad)} &
 \multicolumn{1}{c}{(mrad)} \\ \cline{6-7}
& & & & &  \multicolumn{1}{c}{1-site} &  \multicolumn{1}{c}{2-site}
& \multicolumn{1}{c}{1-site}\\
\hline
\multicolumn{8}{c}{$^{156}$Gd} \\
II & A & $-31$ & $-$40(12) & 1(12) & 50(9) & 56(10) & 52(25) \\
III& A & $-39$ & $-$21(10) & 24(14) & 61(8)& 65(9) & 58(21)\\
IV & B & $-39$ & $-$26(10) & 5(12) & 56(8) & 60(8) & 44(21)\\
\multicolumn{8}{c}{$^{158}$Gd} \\
II & A& $-31$ & $-$38(10) & 35(10) & 68(7) & 77(9) & 102(20)\\
III& A& $-38$ &  $-$55(8) & 59(12) & 97(8) & 105(9) & 155(21)\\
IV & B& $-38$ &  $-$65(8) & 65(10) & 108(8) & 117(9)& 181(21)\\
\multicolumn{8}{c}{$^{160}$Gd$^a$} \\
II &    A & $-30$ &  $-$36(10) &  49(11) & 72(8) & 82(9)& 115(23)\\
III &   A & $-36$ &  $-$64(9) & 48(13) & 96(9) & 104(10) & 157(24)\\
IV &  B   & $-36$ &   $-$84(9) & 64(10) & 117(9) &126(10) &
209(23)\\ \hline \hline \\
%
 \multicolumn{8}{l}{\parbox{5in} {$^a$~A correction has been applied to account for a $\sim 8$\%
 contribution to the intensity of the $4^+ \rightarrow 2^+$ transition in $^{160}$Gd due to the
$9/2^- \rightarrow 7/2^-$ transition in $^{157}$Gd.}}\\ &\\
\end{tabular}
\protect\label{tab:4+results}
\end{table*}

\begin{figure}
    \resizebox{0.8\textwidth}{!}{
  \includegraphics[height=.8\textheight]{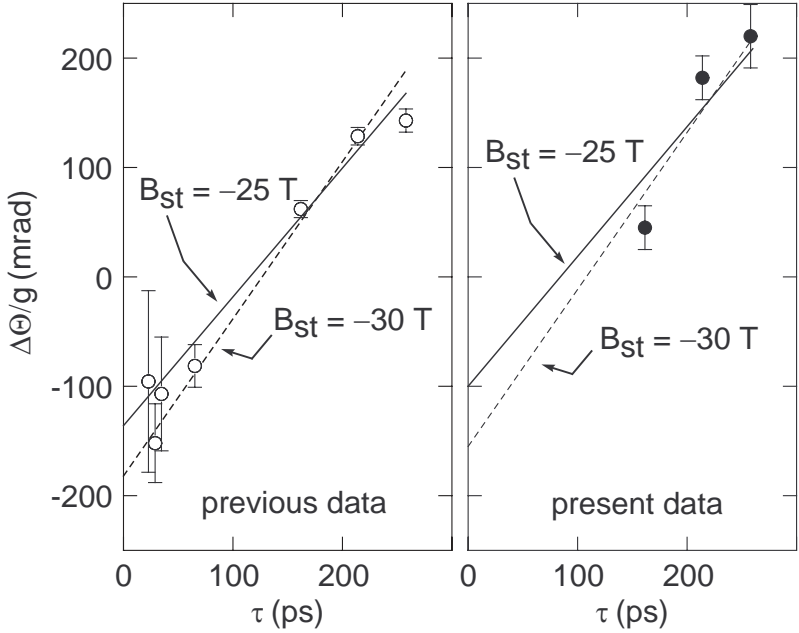}}
\caption{Left: Previous precession results, $\Delta \Theta /g$, for
the 4$^+_1$ and 6$^+_1$ states in the even Gd isotopes \cite{ska72}
plotted as a function of the mean level lifetime $\tau$. These data
were obtained following Coulomb excitation with 56~MeV $^{16}$O
beams. Right: Present data for 4$^+_1$ states in $^{156,158,160}$Gd
from run IV, obtained with 40 MeV $^{16}$O beams. In both panels,
the solid line, for $B_{\rm st} = -25$~T, is the best fit to the
data. The dotted line forces the static field strength to be $-30$~T
as implied by the single-site analysis of the data for the 2$^+_1$
states.}
 \label{fig:phitau}
\end{figure}

If both sides of Eq.~(\ref{eq:DTHETA}) are divided by $g$ it becomes
\begin{equation}\label{eq:DTHETAg}
\Delta \Theta/g = (\omega/g) \tau + \Delta \theta_{\rm tf}/g.
\end{equation}
From this equation it can be seen that a plot of $\Delta \Theta /g$
values versus $\tau$ should lie on a straight line with a slope
$\omega /g$, which is proportional to $B_{\rm st}$, and with an
intercept $\Delta \theta _{\rm tf} /g$ at $\tau = 0$. To aid
comparison with previous work, the final column of
Table~\ref{tab:4+results} shows the total precessions, $\Delta
\Theta$, according to the single-site model. The left panel of
Fig.~\ref{fig:phitau} shows the previous data, from \cite{ska72},
for the 4$^+_1$ and 6$^+_1$ states in the even Gd isotopes so
plotted. For comparison the right panel shows the present data for
the 4$^+_1$ states from run IV (single-site analysis). The results
of the previous work \cite{ska72} have been reanalyzed using the
present adopted $g$~factor and lifetime values (Tables~\ref{tab:tau}
and \ref{tab:g-Q}). These previous data were fitted treating both
$\omega /g$ and $\Delta \theta _{\rm tf} /g$ as free parameters. The
result, shown as the solid line in the left panel of
Fig.~\ref{fig:phitau}, corresponds to $B_{\rm st} = -25(3)$~T, in
agreement with the value of $-25$~T obtained in run IV (single-site
analysis).

Unfortunately the present data for the 4$^+_1$ states are not
sufficiently precise to fit both the slope and the intercept as free
parameters. The transient-field precession was therefore set to the
values adopted in the analysis of the 2$^+$ states (see
Table~\ref{tab:2+results}). These values and the extracted
static-field strengths for the one- and two-site models are
presented in Table~\ref{tab:4+fields}.

\begin{table}
\caption{Hyperfine fields extracted from 4$^+_1$ state precessions.}
\begin{tabular}{ccccc} \hline \hline
 \multicolumn{1}{c}{Run} &
 \multicolumn{1}{c}{Target} &
 \multicolumn{1}{c}{$\Delta \theta_{\rm tf}/g$} &
 \multicolumn{2}{c}{$B_{\rm st}$} \\
 &
 &
 \multicolumn{1}{c}{(mrad)} &
 \multicolumn{2}{c}{(Tesla)} \\ \cline{4-5}
 & & &\multicolumn{1}{c}{1-site} &\multicolumn{1}{c}{2-site} \\
\hline
Ref.~\protect \cite{ska72} & & $-136(26)$ & $-25(3)$\\
II & A & $-80$  & $-16.7(1.3)^a$ & $-19.0(1.4)^a$ \\
III& A & $-93$  & $-22.7(1.4)^a$ & $-24.4(1.5)^a$ \\
IV & B   & $-100$  & $-24.7(1.4)^a$ & $-26.7(1.5)^a$ \\ \hline
\hline &\\
\multicolumn{5}{l}{\parbox{4.0in}{$^a$~The uncertainty in the
transient-field contribution is not included in the quoted error.}}
\\ &\\
\end{tabular}
\protect\label{tab:4+fields} \end{table}

Since the previous measurements \cite{ska72} were made with 56 MeV
$^{16}$O beams, somewhat higher than the 40 MeV beams used here, the
transient-field contribution must be larger in the previous work.
According to the Rutgers \cite{shu80}  and Chalk River \cite{hau83}
parametrizations, the expected difference in $\Delta \theta _{\rm
tf} /g$ is about 30 mrad. As shown in Table~\ref{tab:4+fields}, the
present data for target B in run IV (single-site analysis) are
consistent both with the expected difference in transient-field
precession and with the same static field strength as observed in
Ref.~\cite{ska72}. The present measurements from run IV are also in
agreement with the static-field strengths obtained by H\"ausser {\em
et al.} \cite{hau83}.

All of the previous work has assumed a single-site for the implanted
nuclei. However a comparison of the results in
Tables~\ref{tab:2+fields} and \ref{tab:4+fields} shows that the
assumption of a single implantation site implies that the effective
static-field strengths for the 4$^+_1$ states are significantly
smaller than those experienced by the 2$^+_1$ states.

\subsection{The field-free fraction}\label{sect:fff}

\begin{figure}
    \resizebox{0.8\textwidth}{!}{
  \includegraphics[height=.8\textheight]{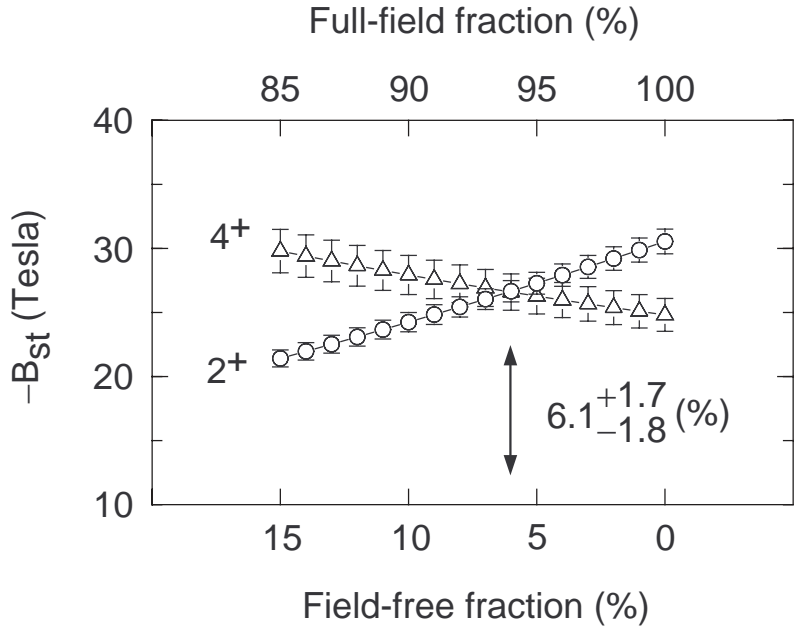}}
\caption{Determination of the field-free fraction for run IV.}
 \label{fig:fff}
\end{figure}

Figure~\ref{fig:fff} illustrates the procedure used to determine the
field-free fraction $f$, according to the two-site model described
in section~\ref{sect:damage}. In order to achieve consistent $B_{\rm
st}$ values for the 2$^+$ and 4$^+$ states requires that in run II,
11.6$^{+2.7}_{-2.3}$\% of the implanted nuclei reside on field-free
sites. In runs III and IV, the field-free fractions are
6.7$^{+1.9}_{-1.8}$\% and 6.1$^{+1.8}_{-1.7}$\%, respectively,
giving an average of $6.4 \pm 1.3$\%. Thus the two-site analysis
adopted $f = 0.116$ for run III and $f =0.064$ for runs III and IV.
The initial expectation would be that $f$ should be the same for all
three runs. It is not clear why a higher percentage of ions
apparently reside on damaged sites in run II. Since the hyperfine
fields experienced by the implanted nuclei in run II are very
different from those in runs III and IV, this difference might
indicate that there is another effect associated with the perturbed
angular correlations that has not been taken into account. The
apparently different value of $f$ was retained for the two-site
analysis of run II, subject to the caveat that it might not
represent the true field-free fraction.

\section{Discussion} \label{sect:disc}

\subsection{Synopsis}

The primary motivation for the present measurements was to obtain an
in-beam measure of the local magnetization of the gadolinium target
foils, which can vary from foil to foil and for different
beam-heating conditions. Clearly, the effective hyperfine field
strength varies with the magnetization. The following discussion
explores the extent to which ratios of effective fields can be
interpreted as ratios of the host magnetizations.

Along with the presentation of the experimental data, the previous
sections have established that the gadolinium foils used in the
present experiments are quasi single crystals in which the
electric-field gradient is directed along the beam direction
(sections~\ref{sect:magmeas} and \ref{sect:ACresults}). It has been
noted that the effect of the electric-field gradient is negligible
for 4$^+_1$ states. It has also been shown that, for the analysis
procedures adopted here, the effect of the electric-field gradient
along the beam direction is essentially negligible for the
longer-lived 2 $^+_1$ states as well. It follows that the presence
of the electric-field gradient does not impede the use of the
effective static-field strength as a measure of the local
magnetization.

Before coming to a discussion of the results in terms of the local
magnetization and beam heating effects (section~\ref{sect:discD}),
it is necessary first to discuss the effect of the hyperfine field
being misaligned with respect to the external field due to domain
misalignment below the saturation magnetization
(section~\ref{sect:discB}).

\subsection{Domain rotation below the saturation magnetization}
\label{sect:discB}

It has been demonstrated for several impurities implanted into iron
that internal hyperfine magnetic fields may be misaligned with
respect to the direction of the external polarizing field, and that
this effect is associated with domain rotation in an incompletely
saturated sample \cite{stu98}. At some level the effect is expected
for all impurity-host combinations.

\begin{figure}
    \resizebox{0.8\textwidth}{!}{
  \includegraphics[height=.8\textheight]{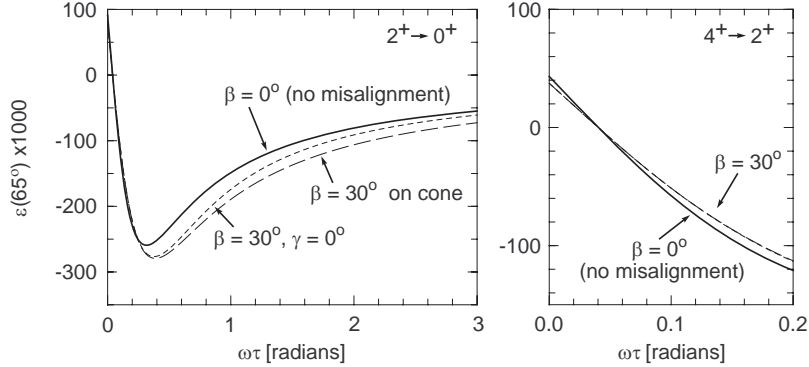}}
\caption{ Asymmetry, $\epsilon(65^\circ)$, as a function of $\omega
\tau$ showing the effect of misalignment between the external field
direction and the internal hyperfine field direction. The angles
$\beta$ and $\gamma$, which specify the direction of the internal
field, are defined as in Fig.~\protect \ref{fig:sketch}. For the
$4^+ \rightarrow 2^+$ case the curves for $\beta = 30^\circ$ on the
cone and $\gamma = 0^\circ$ are almost indistinguishable.}
\label{fig:misaligned}
\end{figure}

Figure \ref{fig:misaligned} shows the effect of misaligned hyperfine
fields on the analysis of the perturbed angular correlations. Two
limiting cases of misaligned fields are assumed. In the first case
the internal fields are assumed to lie on a cone of half angle
$\beta = 30^\circ$, whose axis is the external field direction. In
the second case the internal field is assumed to lie in the plane
defined by the beam axis and the external field direction, i.e. at
angles $\beta = 30^\circ$, $\gamma = 0^\circ$ in terms of the
co-ordinate frame in Fig.~\ref{fig:sketch}. Given that the
gadolinium target has a texture such that it resembles a quasi
single crystal with the $c$~axis along the beam, the latter case is
likely to be more realistic in the present work. The formulae for
evaluating these perturbed angular correlations are given in the
Appendix.

It can be seen from Fig.~\ref{fig:misaligned} that if the internal
field is misaligned with respect to the external field, then the
true precession angle around the internal-field direction is always
larger than that derived when no misalignment is assumed.

At least part of the difference between the effective static fields
observed here and the M\"ossbauer result at 4 K ($B_{\rm st} =
37.3(5)$ T \cite{bau75}) is likely to be associated with
misalignment between the internal and external fields. Indeed the
subtle overall tendency for the effective static fields in
Table~\ref{tab:2+fields} to increase slightly from $^{154}$Gd to
$^{160}$Gd, as the 2$^+_1$ state lifetimes and hence the $\omega
\tau$ values increase, might be associated with misalignment between
the internal and external fields, which has not been included in the
analysis. (Note that the difference between the different curves in
Fig.~\ref{fig:misaligned} decreases as $\omega \tau$ increases.)
However the trend is below the statistical precision of the data.

It can be concluded that although the perturbed angular correlations
may show some sensitivity to the direction of the internal field,
the ratios of $\omega \tau$ values derived assuming that the
internal and external fields are parallel can still be interpreted
as magnetization ratios, to a very good approximation.

\subsection{Beam heating and relative magnetizations of foils}
\label{sect:discD}

There is no significant difference between the {\em ratios} of
effective fields determined from the one-site or two-site analysis
of the data. The following discussion will therefore use the ratios
of the observed precession angles in the simpler single-site
analysis as the measure of the relative magnetizations.
Table~\ref{tab:relmag} summarizes the results of these in-beam
relative magnetization measurements.

\begin{table}
\caption{Relative magnetizations from in-beam precession
measurements.}
\begin{tabular}{lcccccc} \hline \hline
 \multicolumn{1}{l}{Run, Target} &
 \multicolumn{1}{c}{$P_B$  $^a$ } &
 \multicolumn{1}{c}{$D_B$  $^a$} &
\multicolumn{1}{c}{$T$  $^b$ } &
 \multicolumn{3}{c}{Relative magnetization $^c$} \\ \cline{5-7}
  & (W)& (10$^9$ W/m$^3$)&(K) & \multicolumn{1}{c}{$4^+$} &
 \multicolumn{1}{c}{$2^+$} &
 \multicolumn{1}{c}{average}  \\
 \hline
II,  A (Gd) & 0.1   & 1.5  & $ \sim 185$ &  0.57(8) & 0.80(3) & 0.77(3) \\
III, A (Gd) & 0.015 & 0.2  & $ \sim 90$ &  0.83(9) & 0.93(4) & 0.91(4) \\
IV, B (Gd + Cu) & 0.1   & 2.3  & $ \sim 90$ &  1       & 1       & 1
\\ \hline \hline &\\
%
\multicolumn{7}{l}{\parbox{5in} {$^a$~$P_B$ is the power deposited
in the target by the beam. $D_B$ is the power density in the beam spot.}}\\
\multicolumn{7}{l}{\parbox{5in} {$^b$~Calculated temperature at the center of the beam spot.}}\\
\multicolumn{7}{l}{\parbox{5in} {$^c$ Relative magnetizations
determined from ratios of precession angles given in Tables \protect
\ref{tab:2+results} and \protect \ref{tab:4+results}. }}\\ &\\
\end{tabular}
\protect\label{tab:relmag}
\end{table}

The only difference between runs II and III is the beam intensity,
which is nearly an order of magnitude larger in run II (see Table
\ref{tab:expts}). If it is assumed that the local temperature at the
implanted nuclei is near 100 K in run III, the reduction in
magnetization in run II, to $\sim 85$\% of the value near 100 K,
must correspond to a significantly higher local temperature of $\sim
190$~K (see Fig.~\ref{fig:magplot}).

\begin{figure}
    \resizebox{0.8\textwidth}{!}{
  \includegraphics[height=.8\textheight]{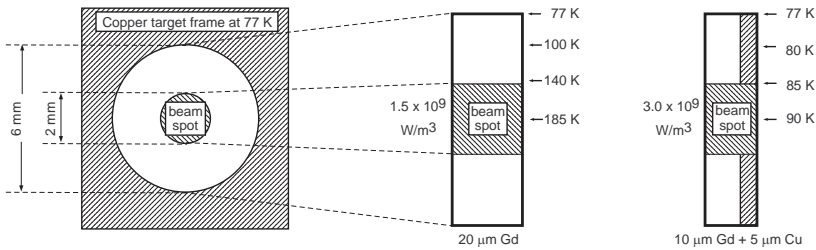}}
\caption{Schematic thermal conductivity calculations. The front view
of the target is shown on the left. To the right are shown side
views of the targets which indicate the temperature profiles between
the center of the target and the point of contact with the target
frame. The two cases shown approximate target A in run II (left) and
target B in run IV (right).} \label{fig:beamspot}
\end{figure}

\begin{figure}
    \resizebox{0.8\textwidth}{!}{
  \includegraphics[height=.8\textheight]{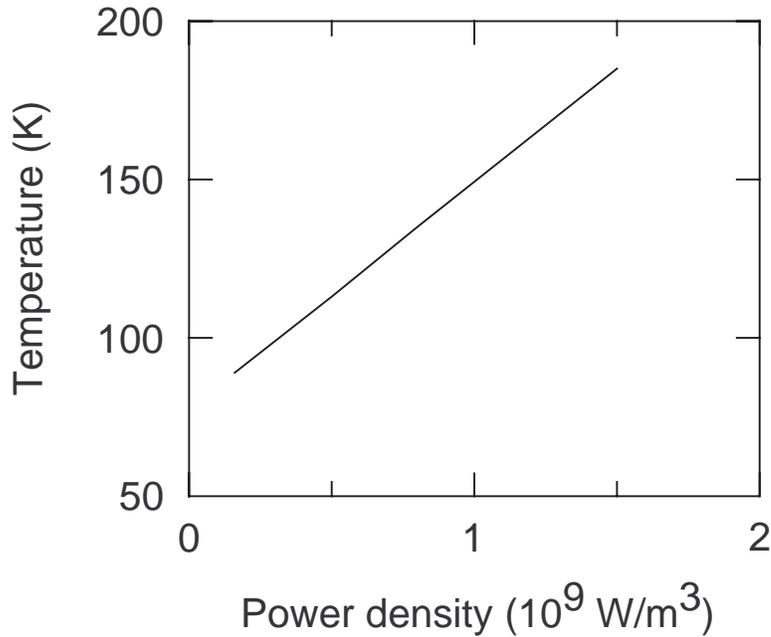}}
\caption{Calculated temperature at the beam spot as a function of
beam power-density, $D_B$, for a 20 $\mu$m thick gadolinium foil.}
\label{fig:beamspot1}
\end{figure}

Thermal conductivity calculations were performed with the QuickField
package \cite{quickfield} to investigate the effects of beam heating
on gadolinium target foils, with and without copper backing layers.
Bulk thermal conductivities were assumed for the gadolinium and
copper layers. Some results are presented in Fig.~\ref{fig:beamspot}
and Fig.~\ref{fig:beamspot1}. The temperature in the beam spot
according to these calculations is included in
Table~\ref{tab:relmag}.

Despite the schematic nature of these calculations, and the fact
that the foils are rather thin (so bulk thermal conductivities may
not apply), the calculations are in very good qualitative agreement
with experiment. The two examples shown in Fig.~\ref{fig:beamspot}
were chosen to resemble target A in run II and target B in run IV.
The schematic calculations reproduce the difference in temperature
required by the experimental data. Fig~\ref{fig:beamspot1} shows the
linear variation in temperature for a 20~$\mu$m thick gadolinium
foil as the beam power is changed. Again, these schematic
calculations reproduce the experimental difference between runs II
and III.

There remains a difference, of the order of 10\%,  between the
magnetizations of target A and target B under conditions where the
beam heating is essentially negligible (runs III and IV). This
difference must be attributed to differences in the crystalline
texture of the foils.

The thermal conductivity calculations also suggest that even a
relatively thin layer of copper evaporated onto a gadolinium foil
will greatly assist in keeping the temperature in the beam spot near
liquid nitrogen temperature, under conditions where the beam
intensity cannot be kept very low. Such a procedure was used for
this purpose in Ref.~\cite{cub93}.

\section{Summary and Conclusions}

The effective hyperfine fields experienced by Gd ions
recoil-implanted into gadolinium foils have been measured for two
targets (one copper-backed) and with differing beam intensities. The
effects of (i) the transient-field interaction, (ii) electric-field
gradients, and (iii) nuclei residing on damaged, field-free sites
have been evaluated. The possible effects of a misalignment between
the external field and the direction of the internal magnetization
was discussed. It was found that the effective hyperfine magnetic
field varies from target to target and with the power deposited by
the beam, particularly when the target is an un-backed gadolinium
foil.

To a very good approximation, the ratios of hyperfine magnetic
fields can be interpreted as ratios of the host magnetization. The
changes in hyperfine field strength, and hence magnetization, with
beam intensity can be correlated with the expected temperature rise
in the beam spot due to the power deposited by the beam. A layer of
copper evaporated onto the gadolinium foil can greatly enhance the
dissipation of beam power and minimize this temperature rise.

The results of the present measurements were used in a recent study
of the transient-fields for high-velocity Ne and Mg ions traversing
gadolinium \cite{stu05}, to correct for differences in the
magnetizations of the gadolinium foils in targets A and B.

\section*{Acknowledgments}
The authors wish to thank Dr N.R. Lobanov for assistance with the
thermal conductivity calculations, and Dr W.D. Hutchison, UNSW@ADFA,
for performing x-ray diffraction measurements on several gadolinium
targets. Dr T. Kib\'edi and Dr P. Nieminen are thanked for their
assistance with the collection of data during run III.

\newpage
\section*{Appendix} \label{sect:appendix}

In this appendix the formulae are given for calculating perturbed
angular correlations in the case where the internal hyperfine fields
are misaligned with the external field. Since a full quantitative
analysis is not required here, it is necessary to consider the case
where only $q=0$ terms are non-zero in Eq.~(\ref{eq:wtheta}). It is
then possible to define
\begin{equation}
a_k = \sqrt{2k+1} \langle \rho_{k0} \rangle F_k Q_k.
\end{equation}

The case where the internal field is assumed to be distributed
equally on a cone of half-angle $\beta$ to the external field has
been considered previously by Ben-Zvi {\em et al.} \cite{ben67}.
Their expression can be rewritten in the form
\begin{eqnarray}
W(\theta_\gamma, \beta) &=& \sum_{k Q q} a_k
[d^k_{q0}(\frac{\pi}{2})]^2
[d^k_{Qq}(\beta)]^2 \nonumber \\
 & \times &
 \frac{\cos(q \theta_\gamma - Q(\Delta \theta_Q + \Delta \theta _{\rm tf}))}
 {\sqrt{1 + Q \omega \tau }} \label{eq:betacone}
\end{eqnarray}
where
\begin{equation}
\tan( Q \Delta \theta_Q) = Q \omega \tau ,
\end{equation}
and for convenience in the present application, the spherical
harmonics in their expression have been replaced by the
$d^k_{Qq}(\beta)$ matrices for the second Euler rotation (about the
$y$ axis).

To our knowledge, the more general case, where the internal field
has a single specific orientation to the external field, specified
by the spherical polar co-ordinates $(\beta, \gamma)$ (see
Fig.~\ref{fig:sketch}), has not been given previously. The result is
\begin{eqnarray}
W(\theta_\gamma, \beta, \gamma) &=& \sum_{k Q q_1 q_2} a_k
 d^k_{q_1 0}(\frac{\pi}{2}) d^k_{q_2 0}(\frac{\pi}{2})
 d^k_{Q q_1}(\beta) d^k_{Q q_2}(\beta)\nonumber \\
 & \times &
 \frac{\cos(q_2 \theta_\gamma -(q_2 - q_1) \gamma -
 Q(\Delta \theta_Q + \Delta \theta _{\rm tf}))}
 {\sqrt{1 + Q \omega \tau }}.
\end{eqnarray}
This expression reduces to Eq.~(\ref{eq:betacone}) when $\gamma$ is
averaged between 0 and 2$\pi$ radians:
\begin{equation}
W(\theta_\gamma, \beta) = \frac{1}{2 \pi} \int_0^{2 \pi}
W(\theta_\gamma, \beta, \gamma) d \gamma
\end{equation}

\end{document}